\documentclass{aa}
\usepackage{epsfig}
\usepackage{graphicx}

\begin{document}

%\thesaurus{11         % A&A Section 4:Galaxies
%              (11.09.3; % intergalactic medium
%	       11.19.6; % Galaxies: structure
%	       10.19.3; % Galaxy: structure
%               11.09.2; % Galaxies: interactions
%	       11.11.1; % Galaxies: kinematic and dynamics
%               11.13.2)} % Galaxies: magnetic fields.

\title{Generation of galactic disc warps due to intergalactic
accretion flows onto the disc}

%\subtitle{.}

\author{M. L\'opez--Corredoira$^{1,2}$, J. Betancort-Rijo$^{2,3}$, J. E.
Beckman$^{2,4}$}
%          \inst{1}

\offprints{martinlc@astro.unibas.ch}

\institute{
$^1$ Astronomisches Institut der Universit\"at Basel, Venusstrasse 7,
CH-4102-Binningen, Switzerland\\
$^2$ Instituto de Astrof\'\i sica de Canarias, E-38200  La 
Laguna, Tenerife, Spain\\
$^3$ Departamento de Astrof\'\i sica, Universidad de La Laguna, Tenerife,
Spain\\
$^4$ Consejo Superior de Investigaciones Cient\'\i ficas (CSIC), Spain
}

\date{Received xxxx; accepted xxxx}

%\markboth{L\'opez-Corredoira et al.}{Warps \& intergalactic flows}

\abstract{A new method is developed to calculate the amplitude of the
galactic warps generated by a torque due to external forces.
This takes into account that the warp is produced as a reorientation
of the different rings which constitute the disc in order to compensate
the differential precession generated by the external force,
yielding a uniform asymptotic precession for all rings.\\
Application of this method to gravitational tidal forces in the
Milky Way due to the Magellanic Clouds leads to a very low amplitude
of the warp, as has been inferred in previous studies; so, tidal forces
are unlikely to generate warps, at least in the Milky Way.
If the force were due to an
extragalactic magnetic field, its intensity would have to
be very high, greater than 1 $\mu $G, to generate the observed warps.\\
An alternative hypothesis is explored: the accretion of the intergalactic
medium over the disk. A cup-shaped distortion is expected, due to
the transmission of the linear momentum; but, this effect is small
and the predominant effect turns out to be
the transmission of angular momentum, i.e.
a torque giving an integral-sign shape warp.
The torque produced by a flow of velocity $\sim 100$ km/s
and baryon density $\sim 10^{-25}$ kg/m$^3$ is enough to generate 
the observed warps and this mechanism offers quite a plausible explanation.
First, because this order of accretion rate
is inferred from other processes observed in the Galaxy, notably 
its chemical evolution. The inferred rate of
infall of matter, $\sim 1$ M$_\odot$/yr, to the Galactic disc that this theory
predicts agrees with the quantitative predictions 
of this chemical evolution resolving key issues, notably
the G-dwarf problem. Second, because the required density of the intergalactic
medium is within the range of values compatible with observation.
By this mechanism, we can explain the warp phenomenon 
in terms of intergalactic
accretion flows onto the disk of the galaxy.
\begin{keywords}
intergalactic medium --- galaxies: structure --- Galaxy: structure ---
galaxies: interactions --- galaxies: kinematic and dynamics --- 
galaxies: magnetic fields
\end{keywords}}

\authorrunning{L\'opez-Corredoira et al.}

\titlerunning{Warps \& intergalactic flows}

\maketitle 

\section{Introduction}
\label{.introd}

Many spiral galaxies present warps, distortions to a flat disc
with an integral-sign shape.
The Milky Way is an example (Burton 1988; 1992). Indeed,
most of the spiral galaxies for which we have relevant information
on their structure (because they are edge on and they are nearby) 
present warps. S\'anchez-Saavedra, Battaner 
\& Florido (1990) and Reshetnikov \& Combes (1998) 
show that nearly half of the spiral galaxies of selected samples are
warped, and many of the rest might also be warped since warps 
in galaxies with low inclination are difficult to detect.

The intergalactic magnetic field has been suggested as the cause of
galactic warps (Battaner et al. 1990;
Battaner et al. 1991; Battaner \& Jim\'enez-Vicente 1998).
This is in our opinion a serious proposal (an opinion not held, however, by
Binney 2000) which could explain many of the observations, although
observational support is still controversial.
The postulated alignment of warps of different galaxies 
(Battaner et al. 1991) and the differences between the gaseous and stellar
warps (Porcel et al. 1997) can have alternative
explanations as we shall see in the present paper.

Gravitational tidal effects on the Milky Way due to the Magellanic Clouds
are not enough to justify the observed amplitude of the warp. Hunter
\& Toomre (1969) calculated that the Clouds with mass $M_{\rm
sat}=10^{10}$ M$_\odot $ and distance $d=55$ kpc would 
generate a warp of amplitude less than 
117 pc at radius 16 kpc in the most favourable case instead of the 
observed 2 or 3 kpc. The Magellanic Clouds are near the pericentres 
of their orbits around Milky Way (Murai \& Fujimoto 1980; Lin \& Lynden-Bell 1982), 
so it is not expected that this amplitude could be greater due to a closer 
approach of the Magellanic Clouds in a recent past.

Weinberg (1998) proposed a mechanism to amplify the tidal effects
due to a satellite by means of an intermediate massive halo around
the galactic disc, but Garc\'\i a-Ruiz et al. (2000)
have found that the orientation of the warp is not compatible with
the generation of warps by means of this mechanism if the satellites
are the Magellanic Clouds. Quantitatively, a better prospect would be
the Sagittarius dwarf galaxy (Ibata \&
Razoumov 1998) since tidal effects are proportional to $\frac{M_{\rm sat}}
{d_{sat}^3}$, the galactocentric distance to this dwarf Galaxy is 
only 16 kpc and its mass $\sim 10^9$ M$_\odot $ (see \S \ref{.gravtor}).

Binney's (1992) review concludes that halos dominate the dynamics
of the warps, although he also points out that {\it ``warps
will in the end prove to be valuable probes of cosmic infall and
galaxy formation''}. In a subsequent paper (Jiang \& Binney 1999),
cosmic infall is used to explain the reorientation of
a massive Galactic halo (8 degrees per Gyr) which produces a warp
in the disc. This model requires a halo ten times more massive than the
disc, an extremely high accretion rate
(3 disc masses in 0.9 Gyr) and, in this scenario, after a sufficiently long
time, the angular momentum of the Galaxy would become parallel
to the direction of the falling matter causing the warp
to decay. This last problem might be solved by including 
a prolate halo (Ideta et al. 2000).
The general case of warps produced by the dynamical friction 
between a misaligned rotating halo and disk was also studied by Debattista 
\& Sellwood (1999). Other proposals which include a massive halo have
also serious flaws or not very plausible assumptions (Nelson \& Tremaine 1995;
Binney et al. 1998).

In spite of the importance which is given to the halo in the dynamics
of galaxies, it is quite possible that they may not even
play a major role in the formation of warps. The mass of the halo of
the Galaxy is not especially well determined and the mass fraction of
galaxy halos might be small (see, 
for instance: Nelson 1988; Battaner et al. 1992; Evans 2001
and references therein).
We should emphasize here that the presence or absence of a very massive
halo will not modify qualitatively the arguments presented below and
would imply quantitative changes within the same order of magnitude. 
We will argue in this paper that warps can be generated
without massive halos or magnetic fields, although our results are
perfectly compatible with the existence of these.

We aim here an alternative solution to those
hitherto hypothesized. The mechanism of generation of warps which is explained
in this paper solves these previous difficulties and does not require 
implausible assumptions.
It is even simpler than the hypotheses previously proposed; it requires only
the infall of a very low density intergalactic medium onto the disc
without the dynamical intervention of an intermediate
halo. It is a very simple idea but it works well, as will be shown below.

An analytical calculation is performed to reduce the
problem to a differential equation and some integrals, which are 
subsequently solved by means of numerical algorithms.
A new method is developed in \S \ref{.warptor}
to calculate the warp parameters from an induced
external torque which takes into account the interaction between
all rings of the galactic disc. The external torque induced by
the accretion is calculated in \S \ref{.toracre}, which allows us to
estimate the required density of the inflow, in \S \ref{.MW}.

\section{Warp induced by an external torque}
\label{.warptor}

In this section, we will explain the mechanism of generation of
warps in a galactic disc due to a net external torque. 
This general method is applicable to any kind of torque acting over
the disc and will be used to derive the properties of the warp
induced by an intergalactic accretion flow.

In order to describe a warp in a galactic disc, 
we will use the usual model of tilted rings (Rogstad et al. 1974): the disc is 
taken to be a set of concentric rings, each of radius between $R_i$ and $R_i+dR$
having angle inclination $\alpha _i$ with respect to the
central disc and intersecting its plane at  
two nodes, perpendicular to the points on the ring
where the elevation is maximum and minimum.
The line joining the nodes (``line of nodes''),
will be taken as common to all the rings, i.e., the nodes for all rings
are aligned, as observed in our Galaxy; this is the equilibrium state
in our model. Therefore, the only parameters which define the warp are:
the direction of the line of nodes and a function, $\alpha (R)$, the
maximum angular elevation of the ring of radius $R$ 
with respect to the plane defined by the central disc.

A torque applied to a rotating rigid body (in our case, the rigid 
body is a ring) produces a precession in it, as
in the case of the equinox precession of the Earth.
The importance of precession in galaxies 
was indeed first recognized by Lynden-Bell (1965).
This section describes how the differential precession in successive
rings generates a warp due to their redistribution.
The general formalism included here is applicable to any kind of 
torque induced by a external force over a set of nearly coplanar rings.
In the following section, the calculation of the torque for the specific case
of the intergalactic accretion flows will be derived.

Other analytical approaches to the problem were used previously.
For instance, the consideration of the warp as a product of equilibrium forces
of vertical components,
such that $F_{\rm ext, z}=F_{\rm grav, z}$, where $F_{\rm grav}$ is the 
gravitational force due to an axisymmetric potential
(used in Kahn \& Woltjer 1959; Binney 1991;
Binney 1992 (his section 2: ``Naive theory''); Battaner \& Jim\'enez-Vicente 1998).
Since the orbit within a ring is tilted, the centrifugal forces will cancel both radial and vertical
components of the gravitational forces so this consideration is not valid. 
Our approach has some advantages and finer calculations than these papers.
A mass in the centre of the galaxy cannot provide any torque to the rings. 
However, the axisymmetric potential of the
disc is distorted by the warp (Binney 1992), and these non-axisymmetries
are those responsible for the warp itself.
The differential orientation of the successive rings rather than a point
mass placed in the centre produces a torque.
The approach used in Hunter \& Toomre (1969) is much better although the
analysis is very different to the one presented here.

\subsection{Equations}

We consider the disc to be made of material on circular orbits with angular
velocity $\omega _{\rm rot}(R)$. If initially all the orbits are in the same plane (the plane
of the disc), perpendicular to a unit vector $\vec{k}$, under the influence of an 
external torque with a non-vanishing component perpendicular to $\vec{k}$ the 
orbits will precess. If the precession is not equal for all orbits they will 
not remain in the same plane, so that the vector perpendicular to the plane 
of a given orbit $\vec{k}(R,t)$ will be a function of $R$ and of time $t$.

In the general case, the component of the external torque along the vector 
$\vec{k}(R,t)$ will induce changes in $\omega _{\rm rot}(R)$.
In addition, other moments of the external force may produce changes in the 
shape of the orbits. However in the applications relevant to the present 
problem these changes are negligible and 
we will take the orbits to remain circular,
and with essentially constant $\omega _{\rm rot}(R)$. 
In this case the dynamics of the disc under an external torque may be reduced
to an equation in partial derivatives for $\vec{k}(R,t)$,
or in other words a set
of an infinite number of ordinary differential equations, (one for each 
value of $R$). The unit vector $\vec{k}(R,t)$ 
is defined by two parameters which are
in fact angles.  

The equations for the evolving system, (throughout this work we 
use equations valid for an inertial, non-rotating frame, and at no stage
do we use a rotating frame) averaged over times longer than the orbital 
periods are:

\begin{equation}
\frac{d\vec{J}}{dt}(R,t)=\vec{\tau}(R,t)=\vec{\tau}_{\rm ext}(R,t)+
\vec{\tau}_{\rm int}(R,t)
\label{juan1}
,\end{equation}
where $\vec{J}(R,t)$ 
is the angular momentum for the ring with radius between
$R$ and $R+dR$, and $\vec{\tau}_{\rm ext}$, $\vec{\tau}_{\rm int}$ 
represent the average torque in the ring due to the
external force, and to the gravitational interaction with the 
rest of the disc respectively.

\subsubsection{External torque}

The external torque is:
 
\begin{equation}
\vec{\tau}_{\rm ext}(R,t)=\tau _{\rm ext}[\vec{k}(R,t).\vec{u},R]\vec{i}_0 
\end{equation}\[
\vec{i}_0\equiv \vec{j}_0\times \vec{k}_0
,\]
where $\vec{u}$ is a unit vector in the direction associated with the cause 
of the external torque, which may be e.g., the direction of the infalling gas 
flow or the position of a perturbing galaxy, $\vec{j}_0$ 
is a unit vector parallel
to the projection of $\vec{u}$ onto the initial plane (perpendicular to
$\vec{k}_0$), 
$\tau _{\rm ext}(x,R)$ is a function of the cosine of the angle 
between $\vec{u}$ and $\vec{k}(R,t)$,
which characterizes the specific mechanism giving rise to the torque.

\subsubsection{Internal torque}

The internal torque is:

\begin{equation}
\vec{\tau}_{\rm int}(R,t)=
\int _{\rm all \ rings}
[d\vec{\tau} _{S}(R,t)dR]dS
,\end{equation}
and $d\tau _{S} (R,t) dR\ dS$ is the gravitational 
torque that the ring of radius between $S$ and $S+dS$ produces
on the ring of radius between $R$ and $R+dR$, which is:

\begin{equation}
d\vec{\tau }_{S}(R,t)dR\ dS=
G\sigma (R) \sigma (S)dR\ dS
\end{equation}
\[
\times \tau _{int}(\vec{k}(R,t).\vec{k(S,t)},R,S)
\vec{k}(R,t)\times \vec{k}(S,t)
,\]
where $\tau _{int}(\cos \alpha _{R, S},R,S)$ is a well
determined function (the same for any mechanism) of the cosine of the 
angle $\alpha _{R, S}$ between the planes of the two orbits. 
Explicitly:

{\small
\begin{equation}
=G\sigma (R)\sigma (S)dR\ dS\ 
\frac{S^2}{R}\int _0^{2\pi }d\phi _1\int _0^{2\pi }d\phi _2
\label{2rings}
\end{equation}\[ \times
\frac{1}
{[1+(S/R)^2-2(S/R)(\sin \phi _1 \sin \phi _2+\cos \phi _1 
\cos \phi _2\cos \alpha _{R, S})]^{3/2}}
\]\[\times
[\sin \phi _1\sin \alpha _{R,S}\cos \phi _2\vec{i}(R,S,t)-
\cos \phi _1\sin \alpha _{R,S}\cos \phi _2\vec{j}(R,S,t)
\]\[
+[\cos \phi _1\sin \phi _2-\cos \phi _2\sin \phi _1
\cos \alpha _{R,S}]\vec{k}(R,t)]
,\]
\[
\alpha _{R, S}\equiv \alpha (S,t)-\alpha (R,t); \ \ \alpha(0,t)\equiv 0;
\]\[
\cos \alpha _{R, S}\equiv \vec{k}(R,t).\vec{k(S,t)}
\]\[
\vec{i}(R,S,t)\equiv \vec{j}(R,S,t)\times \vec{k}(R,t); \ \
\vec{j}(R,s,t)\equiv \frac{\vec{k}(R,t)\times \vec{k}(S,t)}
{|\vec{k}(R,t)\times \vec{k}(S,t)|}
\]
}

The torque between two rings, $d\tau _{S}(R,t)dR\ dS$, is proportional to
$-\vec{j}(R,S,t)$, and the
external torque must be parallel to this. The z-component
of the external torque, if this were present, would produce 
an acceleration in the rotation of the disc rather than warping. 

Other components of a galaxy produce negligible contribution. The bulge 
in practice contribute negligibly to the torque: firstly, 
because it is more spherical than the rings 
and a spherical distribution of
mass produces no torque; and, secondly, because the distance of the bulge
to the outer rings is large enough for these
to produce negligible effects
(the torque is proportional to $S^2/R$ for small $S$, where $S$ is the
radius of the structure).
Numerical experiments were carried out which confirmed this point.
A massive halo, if it exists, would produce an extra internal torque
if the halo is non spherical. In practice, a massive non-spherical
halo will change quantitatively the amplitude of the warp but, qualitatively
the mechanism will be the same; the warp amplitude will be reduced,
never increased, because a massive halo would keep the rings more tightly
bound. At present, no precise calculations including non-spherical
massive halos are given here, but the net result would be to increase
the required accretion rates inferred below.
Only the inner mass of the halo ellipsoid with semimajor axis equal to 
$R$ produce net torque, assuming there is a constant ellipticity halo, whose quadrupole is 
$\sim \frac{4}{15}\frac{v_{\rm rot}^2\epsilon R^3f_h(R)}{G^2}$
($v_{\rm rot}(R)$ is the rotation velocity of the Galaxy; 
$\epsilon $ is the eccentricity of the ellipsoids; a mass distribution for the
halo derived from a hypothetical flat rotation curves due to a dark halo is
assumed; $f_h(R)$ fraction of the mass $M(R)$ embedded in the halo) while 
the disc quadrupole component is $\sim 12\pi \sigma (R_0)e^{R_0/h_R} h_R^4$ 
for a model such as (\ref{sigmaMW}). The disc dominates at $R<22$ kpc 
for $\epsilon=0.2$ and $f_h=0.5$ (Kuijken \& Dubinski 1995), and
the contribution of the halo for $R<16$ Kpc is less than 40\% of the
disc contribution. This means that the order of magnitude of the torque
will be not affected by the inclusion of the massive halo, and will
be perhaps affected by a factor less than $\sim 1.4$ at $R<16$ kpc.

For small angles, in a linear approximation, the proportionality
which results is: 

\begin{equation}
\lim _{\alpha _{R, S}\longrightarrow 0}
d\tau _{S}(R)dR\ dS\propto 
-\alpha _{R, S}\vec{j}(R,S,t)
\label{lim2rings}
.\end{equation}

\subsubsection{Dynamics and evolution of the warp}

The precession 
velocity of $\vec{k}(R,t)$ is much less than $\omega _{\rm rot}(R,t)$,
so we have 

\begin{equation}
\vec{J}(R,t)\approx I \omega_{\rm rot}(R,t)\vec{k}(R,t)
\end{equation}
\begin{equation}
I=2\pi R^3\sigma (R)dR
,\end{equation}
where $I$ is the moment of inertia of the ring,
and $\sigma (R)$ the surface density of the disc.

\begin{equation}
\sigma (R)=\int _{-\infty}^\infty dz \rho _{\rm disc}(R,z)
\label{sigma}
,\end{equation}
where $\rho _{\rm disc}$ is its spatial density, independent of $\phi $ in
an assumed axisymmetric case. 

This approximation is typical in planetary precession, for instance. Using this expression
in (\ref{juan1}), and calling $\vec{\tau}_{\|}$, 
$\vec{\tau}_{\perp}$ the components of $\vec{\tau}$ along $\vec{k}(R,t)$ and 
perpendicular to $\vec{k}(R,t)$ respectively, we obtain

\begin{equation}
|I\dot{\omega }_{\rm rot}(R,t)|=|\vec{\tau} _\| |=|\vec{k}(R,t).\vec{\tau }(R,t)|
\label{Juan3}
;\end{equation}

\[
\frac{d\vec{k}(R,t)}{dt}(R,t)=\frac{\vec{\tau} _\perp}{I\omega _{\rm rot}
(R,t)}
\]\[
=\frac{[\vec{\tau}(R,t)-(\vec{k}(R,t).\vec{\tau }(R,t))\vec{k}(R,t)]}
{I\omega _{\rm rot}(R,t)}
\]

With the set of initial conditions $\vec{k}(R,t_0)=\vec{k}_0\equiv 
(0,0,1)$ $\forall R$,
equations (\ref{Juan3}) 
can be integrated to give the configuration of the orbits, and 
hence the shape of the disc, at any time. 

The method described here is not restricted to linear perturbations 
but is valid as long as the deformed disc is describable by a set of 
tilted rings. However,
for large deformations the orbit cannot remain flat and the approximation 
will fail. 
This is not the case however in the examples given in the present discussion. 
It must be noted that equation (\ref{Juan3}) 
for ($\frac{d\vec{k}}{dt}$) is a first order differential equation,
which implied that it is the angular velocities and not their 
time derivative that are determined by the torque. 
So no matter how large (modulus of vector $\frac{d\vec{k}}{dt}$) may be, 
if the torque vanishes instantly so will $\frac{d\vec{k}}{dt}$. 
This is due to the fact that the energy associated with $\frac{d\vec{k}}{dt}$ 
is much smaller than the energy corresponding to $\omega _{\rm rot}$;
the full equations are obviously second order. 

\subsubsection{Equilibrium configuration}

Here we are interested in 
the stationary, or equilibrium configuration that we assume exists,
rather than in the evolution. It is
clear that in this situation all orbits (or rings) must precess around 
$\vec{u}$ at the same constant speed, $\omega _p$, keeping 
$\vec{k}(R,t).\vec{u}$
constant in time, though different for different orbits.
For $\alpha (R)$ and the precession velocity, $\omega _p(R)$, 
to be independent of time, $\vec{\tau}(R,t)$ must be perpendicular to $\vec{k}(R,t)$ and 
$\vec{u}$ for any $R$. Since this holds for $\vec{\tau}_{\rm ext}
(R,t)$ it must also hold for
$\vec{\tau }_{\rm int}(R,t)$. However,
examining the expression for this last vector in (\ref{Juan3}),
for this condition to hold for all orbits, they must all intersect along
the same straight line, parallel to the unit vector $\vec{i}$. 
In this case the 
position of the orbits is simply given by the function $\alpha(R)\equiv
\cos ^{-1}(\vec{k}(R,t).\vec{k}_0)$.

If the vector $\vec{k}(R,t)$ precesses around $\vec{u}$ with angular velocity
$\omega _p(R)$ we then have:

\begin{equation}
\omega _p(R)\vec{u}\times \vec{k}(R,t)=\frac{d\vec{k}}{dt}
.\end{equation}

From this equation we obtain $\omega _p(R)[\alpha (R)]$ 
as a function of $\alpha (R)$.
Now, since one of the conditions for a stationary configuration is that 
$\omega _p(R)$ be independent of $R$,
setting the derivative of this function with 
respect to $R$ 
equal to zero provides us with the functional equation which 
we must solve for $\alpha(R)$ to give the shape of the distorted disc,

\begin{equation}
\frac{d\omega _p(R)}{dR}[\alpha (R)]=0
\label{domegat}
.\end{equation}

The functional equation (\ref{domegat}) can be solved numerically
using the method explained in appendix \ref{.alpha0}. From this,
we can obtain $\alpha (R)$, i.e. the amplitude of the warp as a function of
the distance from the Galactic centre for a given external torque
$\vec{\tau}_{\rm ext}$.

\subsubsection{Few remarks on the transient regime towards the
equilibrium configuration}

\begin{figure}
\begin{center}
\mbox{\epsfig{file=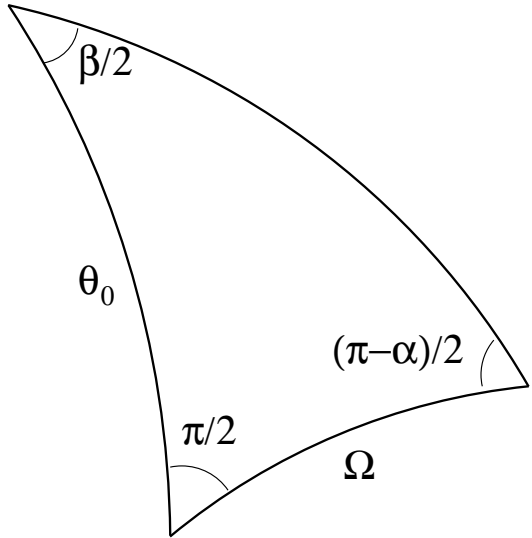,width=7cm}}
%\mbox{\epsfig{file=empty.eps,width=7cm}}
\end{center}
\caption{Spherical triangle which relates different parameter in 
differential precession of two rings.}
\label{Fig:triang2}
\end{figure}

As explained above, we are not concerned here about the transient 
regime. However, it is interesting to comment on the qualitative aspects of 
this regime, since they indicate that some of the mechanisms considered here
may well be of interest. For a mechanism to be acceptable it should lead to
a stationary result, in reasonable agreement with the observations, which are
assumed to correspond to stationary systems. However the mechanism is not
plausible unless it offers a credible scenario for the formation of a warp
once the external influence initiates its operation. We will now see how 
qualitative considerations relative to the warp formation process do 
indicate different degrees of plausibility for different mechanisms, even 
where their predictions for the stationary state are similar.

To carry through this qualitative analysis, it is useful to 
consider eqns. (\ref{Juan3}) 
for a case in which all the mass is concentrated in two
rings (orbits), one of these being much more massive than the other.
$\vec{\tau}_{\rm ext}$
is always perpendicular to $\vec{u}$, 
$\vec{k}$ (although this does not hold for $\vec{\tau}_{\rm int}$),
and since initially $\vec{\tau}_{\rm int}=\vec{0}$ 
the initial motion is a pure precession of 
both rings around $\vec{u}$.
However this precession occurs, in general, at a different
rate for each ring. So the plane of one of the rings will be elevated ``above''
the plane of the other, at a dihedric angle $\alpha $.
After they have precessed
differentially through an angle $\beta $,
and the line of nodes forms an angle
$\Omega $ with the projection of the vector $\vec{u}$ 
onto the plane of the more massive
ring, the relation between these three quantities and $\theta _0$ (where 
$\sin \theta _0= \vec{k}.\vec{u}$) is given implicitly by the 
spherical triangle in Fig. \ref{Fig:triang2}. Clearly as 
$\beta $ goes to zero, $\Omega $ also goes to zero.
This means that the initial node
direction is at $\pi/2$ with respect to the final stationary position.
Now as $\alpha $ increases $\vec{\tau}$ increases, 
producing an additional precession of the plane 
of the lighter ring around the vector perpendicular to the plane of the
heavier.

If the light ring precesses initially around $\vec{u}$ faster than the 
heavy ring (this holds in the general case for rings with an initial 
precession velocity faster than the mean), the additional precession induced
in the heavy ring is in the sense opposed to $\dot{\Omega}$. So this additional 
precession must sweep out an angle somewhat greater than $\pi/2 $ before it 
reaches the stationary state. If the light ring precesses initially more slowly
the corresponding angle would be somewhat less than $\pi/2$. However the 
implication is that the time required to reach the stationary condition is
of order a quarter of the initial precession period. 

Translating this to the
real case, it implies that the time for the formation of the warp must be  
of the order of one quarter of the final global precession period. This time
could be quite long, and a formation procedure which leads to a shorter time
scale would be more plausible. This is the case, for example, where 
$\vec{\tau}_{\rm ext}$ is 
absorbed by the gas. In this case to simulate the evolution we must consider
equation (\ref{Juan3}) 
for two sets of rings, gas rings, which are affected by both 
$\vec{\tau}_{\rm ext}$ and $\vec{\tau}_{\rm int}$,
and stellar rings, which are affected only by $\vec{\tau}_{\rm int}$.
Qualitatively the gas rings will move initially as indicated above,
giving rise to a $\vec{\tau}_{\rm int}$ 
which is perpendicular to $\vec{\tau}_{\rm ext}$. 
The stellar rings which initially 
remain in the same plane start to move under the influence of 
$\vec{\tau}_{\rm int}$, with $\dot{\vec{k}(R,t)}$ parallel 
to $\vec{j}_0$ so that the resulting nodes are already in their final 
position. For very large radii $R$,
the rings would have precesed by a small amount
before the $\vec{\tau }_{\rm int}$ because the torque of the inner stellar
rings, which already are warped, dominate that due to the gas. 
Thus, the line of the nodes of these
orbits will form some angle (going asymptoticaly to $\pi /2$)
with that of the inner orbits. The nodes of the outer orbits trace a
leading spiral. This seems to agree qualitatively with some observations
(Briggs 1990).

The gas rings then move rapidly under the much stronger
$\vec{\tau}_{\rm int}$
generated by the stellar rings, and align generally with them, although the
final warp will be slightly different for the gas rings and the stellar rings,
so that the $\vec{\tau}_{\rm ext}$ experienced by the gas 
is transferred gravitationally 
almost entirely (but not quite) to the stellar ring at the same radius,
although probably the relative displacement between the two is smaller than
the thickness of the disc, and will not be considered in the present
paper.  

As we pointed out earlier, equation (\ref{Juan3}) is first order, so 
if when the orbits arrive at their stationary position 
the torque takes their stationary values (which happens when all 
orbits reach their stationary positions simultaneously) the line of 
the nodes of every orbit will stop precessing in the mean plane defined 
by all other orbits, so stationarity is achieved. However this do not 
seems likely and a more probable outcome is that the line of the nodes 
oscillates around the equilibrium position. Perhaps, friction
could damp the amplitude of these oscillations.
There is no problem in this oscillations being eliminated without 
dissipation (since they contain no energy)
but it seems unlikely because of the degree of conspiracy between 
the orbits required. Thus our stationary solution will correspond 
to the statistical mean, but at any time 
there will be present minor wiggles (in $\alpha(R)$) superimposed
on it.

The comment on warp formation presented here is a qualitative 
anticipation of future detailed work and does not affect any 
of the conclusions of the present work.

\subsection{Example of application: gravitational torque}
\label{.gravtor}

Once we have a general formalism in which a warp is generated when
a torque is produced in a galactic disc, we can analyze possible mechanisms
to produce that torque. A conventional example is the torque generated by
gravitational tidal effects.

\begin{figure}
\begin{center}
\mbox{\epsfig{file=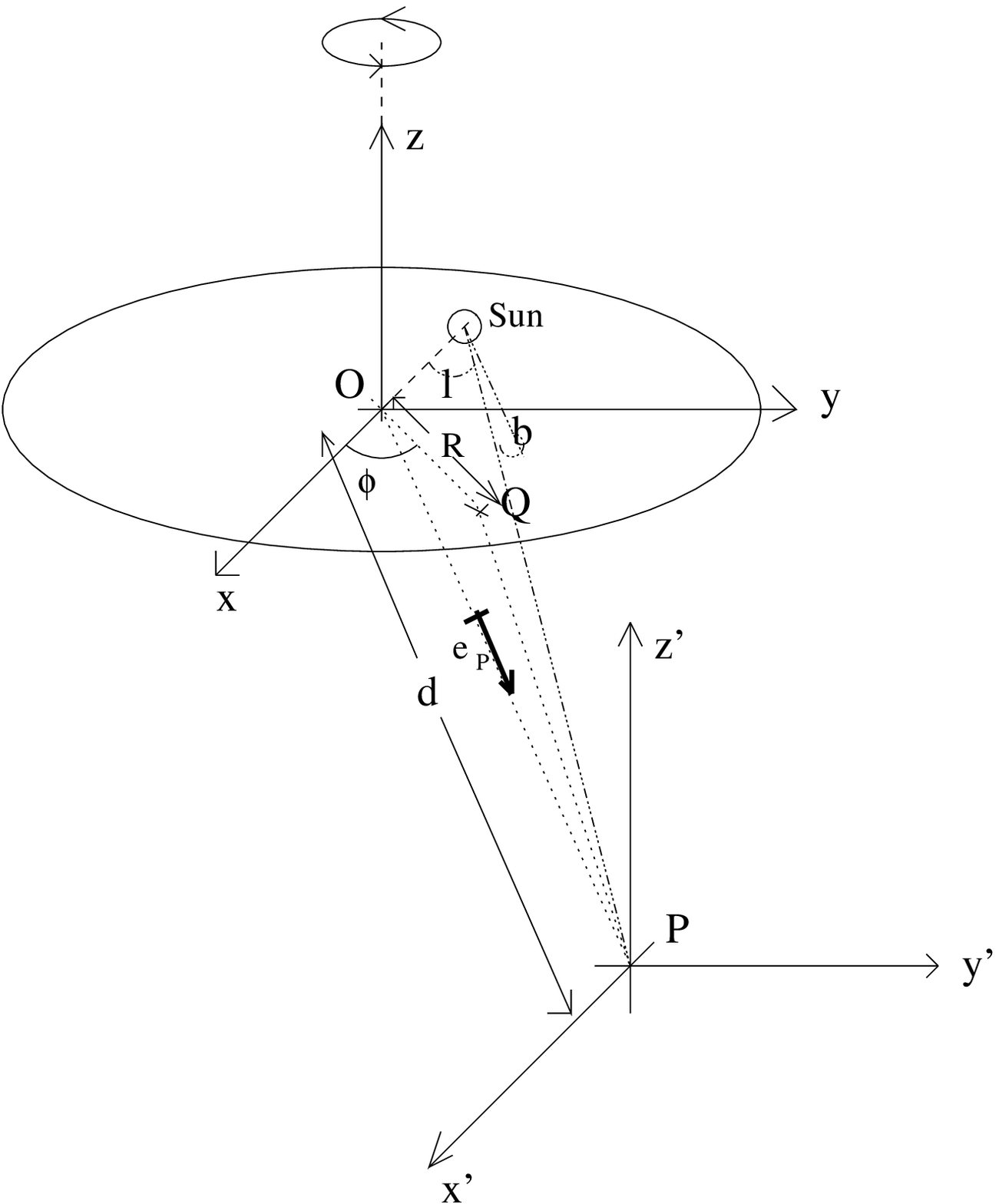,width=8cm}}
%\mbox{\epsfig{file=empty.eps,width=8cm}}
\end{center}
\caption{Graphical representation of the Galactic disc (with coordinates
xyz) and a point mass at $P$ (with coordinates x'y'z').}
\label{Fig:disc}
\end{figure}

In Fig. \ref{Fig:disc}, we show the
Galactic disc centred at $O$ and a point mass at $P$ such that
$\overline{OP}=d\vec{e_P}$. In the case of the Milky Way disc, the Sun would
be situated in the disc with $R=R_0$, $\phi =180^\circ $.
The unit vector can be expressed as

\begin{equation}
\vec{e_P}=\cos \phi_P\cos \theta _P\vec{i}+
\sin \phi_P\cos \theta _P\vec{j}+\sin \theta _P\vec{k}
\label{eP}
.\end{equation} 

It is well-known that the torque produced by a point mass
of mass $m_P$ on any axisymmetric body is, considering only the
quadrupolar term,

\begin{equation}
\vec{\tau _{\rm grav}}\approx \frac{3G\ m_P(I_3-I_1)}{2d^3}\cos \theta _P
\sin \theta _P 
\label{gravtorque}
\end{equation}
\[\times
(\sin \phi _P \vec{i}-\cos \phi _P \vec{j})
,\]
where $I_1$ and $I_3$ are the inertia tensor components (in this
case for the disc).

The solution of (\ref{domegat}) with $\tau _{\rm ext}=\tau _{\rm grav}$
by means of the numerical method explained
in appendix \ref{.alpha0} gives $\alpha (R)$ for the galactic warp.
It will depend on the adopted Galactic model.
For the Milky Way, a suitable model can be adopted as follows:

The surface density of the disc, (\ref{sigma}), is taken as

\begin{equation}
\sigma (R)=
\left \{ \begin{array}{ll} 
        \sigma (R_0)e^{-\frac{R-R_0}{h_R}} ,& 
	\mbox{ $R \le 3R_0$} \\ 
0,& \mbox{ $R > 3R_0$} 
\end{array} \right \}
\label{sigmaMW} 
;\end{equation}
where the local surface density is 
$\sigma (R_0)=48$ M$_\odot $pc$^{-2}$
(Kuijken \& Gilmore 1989), 
the distance to the Galactic centre is $R_0=7.9$ kpc 
(L\'opez-Corredoira et al. 2000);
and the scale length $h_R=3.5$ kpc (Bahcall \& Soneira 1980).
We truncate the exponential disc
at $3R_0$. The Galactic disc undoubtedly extends to larger radii but the effects of
those outer rings can be considered negligible in the present dynamical
context. 

The rotation velocity is taken as

\begin{equation}
v_{rot}(R)=R\omega _{rot}(R)
\label{rotvelMW}
\end{equation}
\[=
\left \{ \begin{array}{ll} 
        200\ {\rm km/s} ,& \mbox{ $R \le 15 \ {\rm kpc}$} \\ 
200\sqrt{15\ {\rm kpc}/R}\ {\rm km/s},& \mbox{ $R>15\ {\rm kpc}$}
\end{array} \right \}
.\]
(Honma \& Sofue 1996). Outside the stellar truncation radius ($\sim 15$ kpc), 
a Keplerian law ($v_{\rm rot}\propto
R^{-1/2}$) is followed, which implies that we have neglected
dark matter contributions from larger radii (according to Honma \& Sofue 1996,
there is no essentialy dark matter beyond 15 kpc).

For the Large Magellanic Cloud, 
we adopt $\theta _P=-33^\circ $, $m_P=10^{10}$
M$_\odot $ and $d=55$ kpc, the same values that Hunter \& Toomre (1969)
adopted. 

\begin{figure}
\begin{center}
\mbox{\epsfig{file=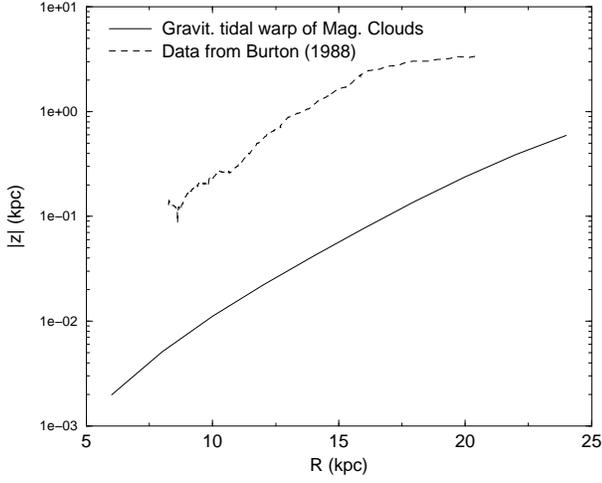,width=8cm}}
%\mbox{\epsfig{file=empty.eps,width=8cm}}
\end{center}
\caption{Milky Way warp maximum height 
as a function of the radius due to
gravitational tidal effects of the Magellanic Clouds (solid line). 
Dashed line stands for observational data of the northern warp
(Burton 1988) rescaled to $R_0=7.9$ kpc.}
\label{Fig:alfagrav}
\end{figure}

The precession we obtain is $\omega _p=2.7\times 10^{-19}$ rad/s
and the amplitude of the warp is shown in Figure \ref{Fig:alfagrav} where
$\alpha (R)\approx |z|/R$. The functional shape of $\alpha (R)$ resembles that
of observational data, but the amplitude is very different: a factor of
20 or 30 separates the two curves, in agreement with Hunter \& Toomre (1969).
This means that we would require the mass of
the Magellanic Clouds to be around 2 or $3\times 10^{11}$ M$_\odot$
to justify the Milky Way warp as a product of the gravitational interaction
with the Clouds. The calculation
of the amplitude with the Saggitarius dwarf galaxy ($m_P=10^9$ M$_\odot $,
$d=16$ kpc; Ibata \& Razoumov 1998) 
gives an amplitude 4 or 5 times larger, which is still not
enough to produce the warp, nor is the predicted 
direction of the warp in agreement with the observations. 

\subsection{Example of application: magnetic torque}

Another example can be calculated: the case of magnetic forces
(Battaner et al. 1990;
Battaner \& Jim\'enez-Vicente 1998), in which the force per unit volume
which produces the torque for a $\theta _P=45^\circ $ inclination field is:

\begin{equation}
\vec{F}=\frac{B^2\sin (2(\theta _P-\alpha (R))}{16\pi L}\sin(\phi -\phi _P)\vec{k}
\label{fmag}
,\end{equation}
where $L=1$ kpc is the adopted value for the characteristic length
in which galactic regions dominated by the galactic magnetic field
become dominated by the extragalactic magnetic field.
Hence, the torque over a ring between $R$ and $R+dR$ is:

\begin{equation}
\vec{\tau }dR=2h_zR\ dR\int _0^{2\pi }d\phi \ \vec{r}\times \vec{F}
\end{equation}\[=
\frac{B^2R^2h_z\sin (2(\theta _P -\alpha(R)))dR}{8L}(\sin \phi _P\vec{i}-\cos \phi _P
\vec{j})
,\]
where $h_z=0.1$ kpc is the scale height of the disc.

With the same Galactic model as the previous subsection, we obtain
that the amplitude of the magnetic field is

\begin{equation}
B\sim 1.4\ {\rm \mu G}
,\end{equation}
in order to obtain a coincidence with observational Burton (1988) data.
The precession angular velocity is $\omega _p=5.1\times 10^{-18}$ rad/s.

\begin{figure}
\begin{center}
\mbox{\epsfig{file=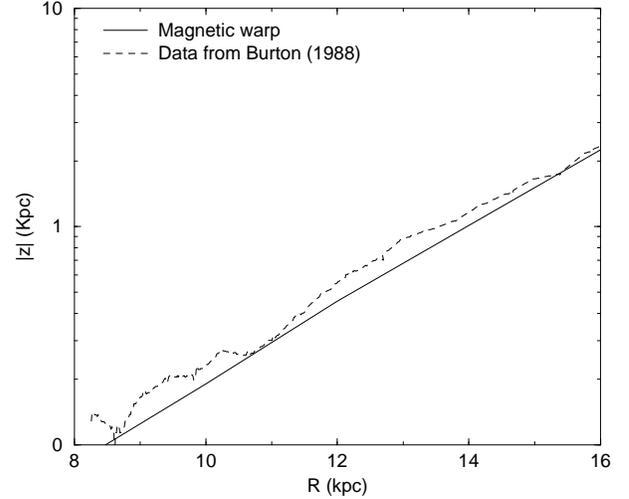,width=8cm}}
%\mbox{\epsfig{file=empty.eps,width=8cm}}
\end{center}
\caption{Milky Way warp maximum height 
as a function of the radius due to
an extragalactic magnetic field of intensity 1.4 $\mu $G (solid line). 
Dashed line stands for observational data of the northern warp
(Burton 1988) rescaled to $R_0=7.9$ kpc.}
\label{Fig:alfamag}
\end{figure}

There is a good agreement for the curves in Fig. \ref{Fig:alfamag},
which was noted by Battaner \& Jim\'enez-Vicente (1998).
The intensity of the field required
to produce the warp is perhaps somewhat high, but 
we will not discuss in this paper
the possibility of the existence of such a field. 
Kronberg (1994), for instance, argues that the value of
of the intergalactic magnetic field can be as high as this.
We remark only that the possibility of
warps generated by extragalactic magnetic fields should not be taken
lightly, although the high value on the required field is perhaps
questionable.
This value is in agreement with that of Battaner \& Jim\'enez-Vicente (1998)
although their calculation method is much simpler.
Binney (1991) obtained a higher required value for $B$, with a difference
of an order of magnitude, in part due to his adoption of an expression for
the magnetic force different from 
expression (\ref{fmag}) equivalent to a much lower
value of $L$. He also adopted an insufficiently precise 
approximation of an axisymmetric potential not distorted by the warp.

\section{Generation of warps by an intergalactic flow}
\label{.toracre}

\subsection{Accretion of matter onto the galactic disc and warps}

The idea we are suggesting here is not totally new. It is based on
ideas about infall of intergalactic matter 
previously explored by other authors
(Kahn \& Woltjer 1959; Binney \& May 1986;
Ostriker \& Binney 1989; Jiang \& Binney 1999).
Kahn \& Woltjer (1959) first suggested that an intergalactic matter flow could
bring about the warp; however, their rough calculations
in fact derive a pressure gradient transmitted to the disc
by means of a hypothetical halo compressed by a subsonic massive wind
and do not specify the mechanism of generation of the warp.
The representation they use is very simple, and quite different from
that presented here. The dominant response to the intergalactic medium
ram pressure of a disc moving through it would be axisymmetric, taking
the form of a rim rather than a warp (Binney 1992),
although Kahn \& Woltjer's estimates of the amplitude in fact agree with our
values and their work was a first indication of a possible
mechanism to explain warps.
Ostriker \& Binney's (1989) suggestion of how cosmic infall can bring 
about warped galactic discs is rather qualitative and they did not proceed 
to a quantitative analysis. 
Subsequent papers (Binney et al. 1998; Jiang \& Binney 1999) 
use a model based on the infall onto
the Galactic halo and do not consider the direct infall onto the disc.
The latter is the novelty here: we take the idea of cosmic infall, and explore
the torque it creates when it collides transmitting
its angular momentum, but we consider directly the interaction
with the disc. We do not invoke the idea of a massive halo modulating the
dynamical effects on the disc.

The idea of cosmic infall has been considered in the context of CDM theory.
Ryden \& Gunn (1987) and Ryden (1988) have shown 
that half of the total angular momentum of any galaxy
was contributed by material that fell in over the last third of a Hubble time.
Galaxy formation theories require this infall. 
Furthermore, many observations imply that there must be an
infall of material to galaxies (Binney 2000):
Local Group members approach each other, the high velocity clouds
around the Galaxy have on the average a net negative velocity (Blitz et al.
1999; Braun \& Burton 1999), and others.

There are good reasons to believe that the infalling baryonic matter
is accreted directly by the disc rather than the halo
(non-baryonic matter can escape completely or be captured by the halo).
The principal supporting arguments here are those based on 
the observed chemical evolution (Ostriker \& Binney 1989; 
L\'opez-Corredoira et al. 1999). 
Significant accretion of metal poor gas is necessary to justify the observations concerning
star formation and metallicity distribution in the galactic disc,
often termed the G-dwarf problem (Tinsley 1980) and the details of
time-dependent evolution of individual metals (Casuso \& Beckman 1997; 2000). 
Recent results implying that this accretion has been constant, or even increased, during
the disc lifetime, are found in Rocha-Pinto et al. (2000).
Moreover, it is clear that a halo should trap accreted
matter with low efficiency, since its mean baryonic density is very low.

In a more general context, it should not be thought that the general secular 
infall of matter is the only factor. Any cloud in the intergalactic medium 
whose orbit intersects the galaxy and is accreted by the disc provides a 
torque due to the interchange of its angular momentum. 
For instance, the exchange of matter between two
galaxies can supply accretable intergalactic matter: 
an intergalactic flow is produced between the two galaxies. 
The HVCs (High Velocity Clouds) have been suggested 
(Blitz et al. 1999; L\'opez-Corredoira et al. 1999; 
Wakker et al. 1999a; Binney 2000) to be observable evidence of the
material which is continuously falling onto the Galactic disc.

\subsection{Description of the flow}

An intergalactic flow can be described as 
a beam of particles which comes from infinite distance towards the galactic
disc with velocity $\vec{v_0}$. Each particle of the beam follows
a trajectory which is not a straight line, due to the gravitational
attraction of the galaxy, until it reaches the galactic plane ($z=0$).
As it intersects the plane, it collides with the gas of the galaxy
and remains trapped in the disc. 
A torque results from the angular momentum contribution of
this particle to the disc. The net torque over a ring of the disc of radius
between $R$ and $R+dR$ will be the sum of the torques produced by all 
the particles of the beam which collide with the galactic disc at distance
between $R$ and $R+dR$ from its centre. 

The total angular momentum with respect to the centre of the galaxy
which is transported by a cylindrical beam of these characteristics with axis
crossing the centre of the 
galaxy is zero. Is the total angular momentum deposited in each ring also zero?
The answer to this question is ``no'', and this is
the key to warp generation by an intergalactic flow.
The net momentum transferred to each ring is non-zero 
because, for a general case where the net flow is not perpendicular
to the plane of the galaxy nor isotropic, the particles which fall
onto a given ring do not come from a single cylindrical shell of the flow. 
The particles are redistributed, and so is their angular momentum, due to the gravitational
interaction with the galaxy. Therefore, the impact parameter of each infalling
particle is not the same for all the particles which cut a given ring (if
it were the same the net contribution of the angular momentum would indeed 
be zero); this will be analytically expressed in the equations given below in
the present section. In particular,
the variation of the impact parameter for a ring as a function of
azimuthal angle is expressed in eq. (\ref{RQ}). 

\begin{figure}
\begin{center}
\mbox{\epsfig{file=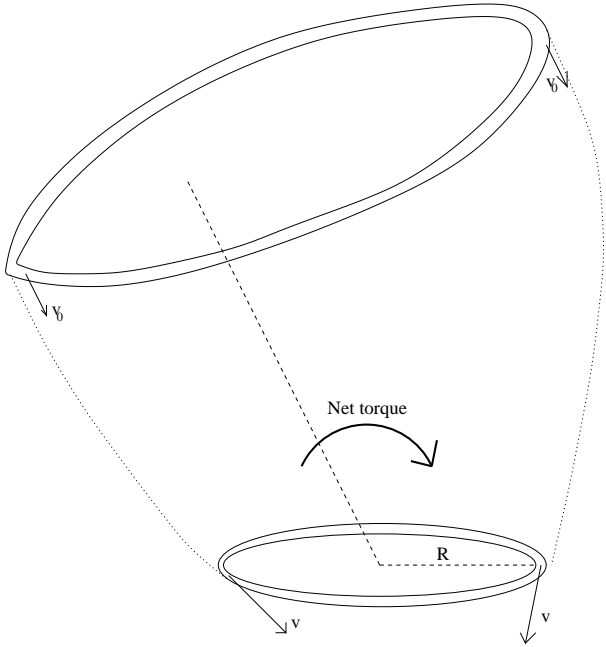,width=8cm}}
%\mbox{\epsfig{file=empty.eps,width=8cm}}
\end{center}
\caption{Graphical representation of the infall of an intergalactic
flow (non-circular ring, above) which produces a net torque on a circular ring in the galactic disc (below).}
\label{Fig:infall}
\end{figure}

The calculations in subsections below are, perhaps, a little complicated
to follow. However, the qualitative
description of the physical system whose variables we will calculate
is not difficult to understand. Figure \ref{Fig:infall} gives a pictorial
description which should be helpful. There is a net torque because the
set of particles which fall into the ring between radii $R$ and $R+dR$
comes from a non-circular ring (which is not in fact elliptical). 
The transformation of this non-circular ring to the circular ring in the disc plane
is effected by gravity. The velocity of
the impacting particles varies with azimuth in the galactic plane ring. 
We wish to calculate the total angular
momentum which is transported by matter in the non-circular ring which falls
into the circular ring. Conservation of angular momentum implies
that the angular momentum transmitted to the circular
ring must be the same as that in the non-circular ring.
Nevertheless, calculation of this angular momentum is not easy
because the geometrical shape of the non-circular ring is not
easy to describe analyticaly. This gives rise to the rather 
tedious calculations in subsections below.

Once a torque is produced over each ring
a warp is generated to compensate the differential precession of the rings,
as we have seen in the previous section. 
We have explained the mechanism of
generation of warps when any kind of torque due to external forces is applied
over the disc, in \S \ref{.warptor}. Now, in this section, we
carry out the calculation of the torque due to the intergalactic flow.

\subsection{Torque due to collision with a particle}
\label{.1part}

\begin{figure}
\begin{center}
\mbox{\epsfig{file=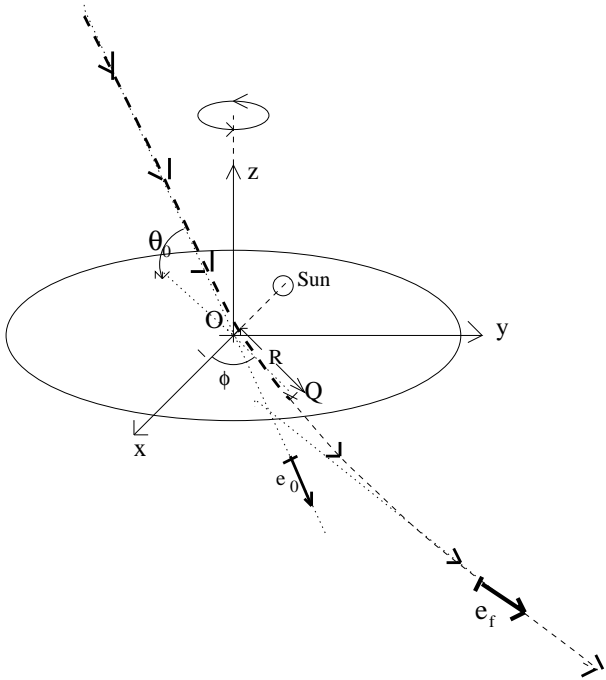,width=8cm}}
%\mbox{\epsfig{file=empty.eps,width=8cm}}
\end{center}
\caption{Graphical representation of the Galactic disc in the coordinates
system xyz, and the hyperbolical trajectory of an infall particle.}
\label{Fig:disc2}
\end{figure}

In Fig. \ref{Fig:disc2}, we give a graphical representation of 
the galactic disc centred at $O$, and the trajectory of a particle
which intersects the disc. 
A particle comes from infinite distance
with velocity $v_0$ in the normalized direction $\vec{e_0}$ given by
angles $\phi _0$, $\theta _0$ in spherical coordinates  
(in Fig. \ref{Fig:disc2}, $\theta _0<0$), where

\begin{equation}
\vec{v_0}=v_0(\cos \phi_0\cos \theta _0\vec{i}+
\sin \phi_0\cos \theta _0\vec{j}+\sin \theta _0\vec{k})
,\end{equation}
and it follows a trajectory which crosses the Galactic
disc in some point $Q$ whose distance from the centre is $R$
and angle with respect to the $x$-axis (defined as the line
``Galactic centre-Sun'' with $x$ negative towards the Sun) is $\phi $,
i.e.

\begin{equation}
\overline{OQ}=\vec{R}=R\cos \phi\ \vec{i}+ R_Q\sin \phi\ \vec{j}
.\end{equation}

A minor order correction for small warp amplitudes, 
that we also take into account, is
the variation of angle of the flow in
the warped rings. We must bear in mind that

\begin{equation}
\theta _0(R)=\theta _0(R=0)-\alpha (R)
.\end{equation}

\begin{figure}
\begin{center}
\mbox{\epsfig{file=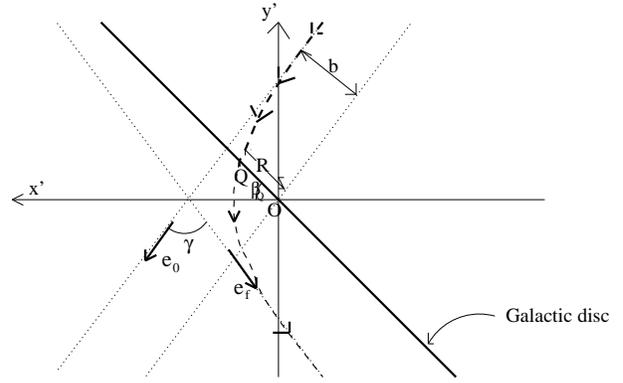,width=8cm}}
%\mbox{\epsfig{file=empty.eps,width=8cm}}
\end{center}
\caption{Graphical representation of the hyperbolical trajectory of an 
infall particle in the plane of the orbit.}
\label{Fig:orbit}
\end{figure}

The same trajectory is represented in Fig. \ref{Fig:orbit}
in the plane of the orbit $x'y'$. It is assumed that the trajectory
is a hyperbola typical of a two-body gravitating system where
the heavier body is the galaxy whose mass is $M_{\rm gal}$ concentrated
at the point $O$. Some minor effects due to the dispersion of the mass
throughout the disc are expected but they are negligible if $R$ is larger
than several disc scale lengths (i.e. greater than $R\approx 10$ kpc).
The orbit is a hyperbola because the
energy of the system is positive, since the velocity at infinite distance,
$|\vec{v_0}|$, is greater than zero. Therefore, the plane of the orbit
will be determined by the independent vectors $\vec{r}$ and $\vec{v_0}$
and the equation of the orbit in the $x'y'$ system is ($r$ and $\beta $
are the polar coordinates in this system; see Fig. \ref{Fig:orbit}):

\begin{equation}
\frac{\epsilon}{A\ r}=1+\epsilon \cos \beta
\label{orbit}
,\end{equation}
where $\epsilon $ is the eccentricity of the orbit,

\begin{equation}
\epsilon =\sqrt {1+\left(\frac{b\ v_0^2}{G\ M_{\rm gal}}\right)^2}
\label{epsilon}
,\end{equation}
and

\begin{equation}
A=\sqrt {\left(\frac{G\ M_{\rm gal}}{v_0^2b^2}\right)^2+
\frac{1}{b^2}}=\frac{v_0^2}{GM_{\rm gal}}\frac{\epsilon}
{\epsilon ^2-1}
\label{A}
.\end{equation}

The impact parameter is $b$ and the net asymptotic angular deviation
$\gamma $ (see Fig. \ref{Fig:orbit}) is given by

\begin{equation}
\tan \frac{\gamma }{2}=\frac{1}{\sqrt{\epsilon ^2
-1}}=\frac{GM_{\rm gal}}{bv_0^2}
\label{gamma}
.\end{equation}

The determination of the point of intersection 
of the orbit with the disc of the galaxy
is a simple trigonometric problem. From a triangle shown Fig. \ref{Fig:orbit},
we can derive:

\begin{equation}
\beta _Q=\frac{\pi}{2}+\frac{\gamma }{2}-\cos^{-1}(e_{0Q})
,\end{equation}
\begin{equation}
e_{0Q}=\cos(\vec{v_0}, \vec{r_Q})=\frac{\vec{v_0}\vec{r_Q}}{v_0r_Q}=
\cos (\theta _0)\cos (\phi _0-\phi )
\label{e0Q}
.\end{equation}

From these expressions, together with (\ref{orbit}), (\ref{epsilon}),
(\ref{A}) and (\ref{gamma}), the radial galactocentric
distance $R$ of the point of orbit intersection with the Galactic plane is:

\begin{equation}
R=r_Q=\frac{b^2v_0^2}{b\ v_0^2\sqrt{1-e_{0Q}^2}+G\ M_{\rm gal}(1-e_{0Q})}
\label{RQ}
.\end{equation}

The angular momentum of the particle with mass $dm$
is constant along its trajectory, 

\begin{equation}
\vec{J}=dm\ v_0 b\frac{\vec{r_Q}\times \vec{v_0}}
{r_Q \ v_0 |\sin(\vec{v_0}, \vec{r_Q})|}
,\end{equation}
and the torque produced by a particle which transmits its angular
momentum to the disc is:

\[
\vec{\tau }=\frac{d\vec{J}}{dt}=\frac{dm}{dt}v_0b\frac{\vec{r_Q}\times \vec{v_0}}
{r_Q\ v_0\sqrt{1-e_{0Q}^2}}=\frac{dm}{dt}v_0b\ (1-e_{0Q}^2)^{-1/2}
\]\begin{equation}\times
[\sin \phi
\sin \theta _0\vec{i}-\cos \phi \sin \theta _0 \vec{j}+
\cos \theta _0\sin (\phi _0-\phi)\vec{k}]
\label{tau1}
.\end{equation}

The new material is stopped by the friction with the disc and its
angular momentum added to the ring. Note that although 
the ring is not really a
rigid body it behaves like one. A single particle in orbit around the
Galactic centre can resemble the dynamics of the rigid body;
it carries the increase of angular momentum and the orbit is distorted
according to the applied torque.

The flow is stopped by the friction with the gas, so the angular momentum
is, at first, transmitted to the gas. This does not mean that gas disc
warps while stellar disc does not. Indeed, it is expected that stars in
the ring feel the gravitational torque due to the gas rings and are
dragged by them. There may be some difference between the stellar warp and
the gas warp due to this lag in the dynamics, but the
difference is likely to be small.
If the stellar disc were
demonstrated to be less warped than the gas warp, it would be 
evidence in favour of either this theory or the theory
of the intergalactic magnetic field as the generator of the warp
(Porcel et al. 1997), or indeed of any theory in
which the external torque affects directly the gas disc rather than
the stellar disc.

\subsection{Total torque due to collision with a particle beam}
\label{.torque}

\begin{figure}
\begin{center}
\mbox{\epsfig{file=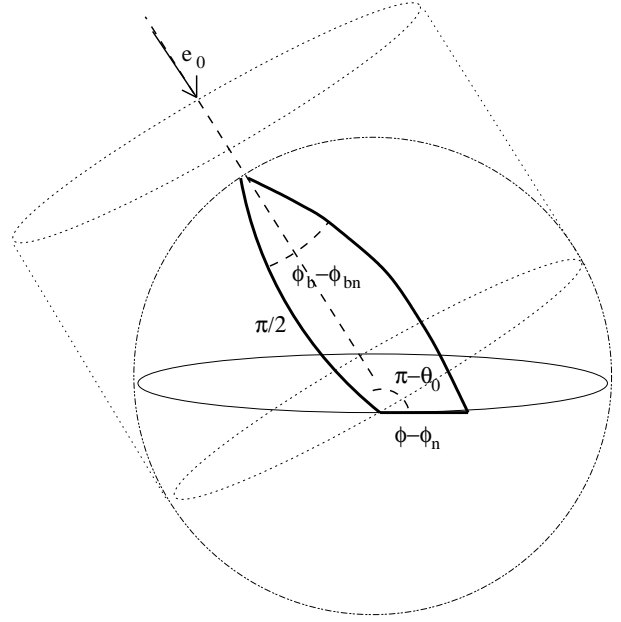,width=8cm}}
%\mbox{\epsfig{file=empty.eps,width=8cm}}
\end{center}
\caption{How the polar angles $\phi _b$ in the original flow and 
$\phi $ in the plane of the galaxy are related in terms
of a spherical triangle.}
\label{Fig:planeperp}
\end{figure}

Expression (\ref{tau1}) gives the torque produced by a particle
which falls to the disc with an impact parameter $b$ and intersects the
disc at $\vec{r_Q}$. If we want to know the total torque produced by
all the particles which come with any $b$ and intersect the disc at
a distance between $R$ and $R+dR$ with any azimuth $\phi $, 
we have to integrate over all
the particles of the beam which fall in this ring. 
The whole beam is then represented
by the varying in the plane perpendicular
to $\vec{v_0}$ the initial (at infinite distance) position of the
falling particle, whose polar coordinates are $b$ and $\phi _b$
(see Fig. \ref{Fig:planeperp}). Thus, the total torque exerted over
the ring with radius between $R$ and $R+dR$ is

\[
\vec{\tau}(R)dR=\int _0^{2\pi }d\phi _b\int _{0; R<R_Q<R+dR}^\infty db b
\]\[ \times
[\frac{dm}{dt}v_0b\ (1-e_{0Q}(\phi )^2)^{-1/2}[\sin \phi
\sin \theta _0\vec{i}
\]\begin{equation}
-\cos \phi \sin \theta _0 \vec{j}+
\cos \theta _0\sin (\phi _0-\phi)\vec{k}]]
\label{tau2}
,\end{equation}
\begin{equation}
dm=\rho _{\rm b}\ v_0dt
,\end{equation}
where $\rho _{\rm b}$ is the density of baryonic matter in
the particle beam, assumed to be independent
of $b$ and $\phi _b$. 
Any non-baryonic matter in the inflow would
not be trapped in the disc so it should not be taken into account within
the total mass of the flow for the purpose of computing the torque.
In the notation, $\vec{\tau }$ stands for
the torque per unit galactocentric radial length.
Note that $\phi _b$ is the polar angle in the 
plane perpendicular to $\vec{v_0}$ and is different from 
the polar angle $\phi $ in the Galactic disc. The relationship between 
the two is:

\begin{equation}
\cot (\phi _b-\phi_{bn})=\frac{\cot (\phi -\phi _n)}{|\sin \theta _0|}
,\end{equation}
according to the general formula relating the angles in the
spherical triangle of Fig. \ref{Fig:planeperp}. $\phi _n$ and $\phi _{bn}$
are the polar angles of the galactic disc and the plane perpendicular
to $\vec{v_0}$ respectively corresponding to the node where the two planes
intersect. We can choose for convenience the origin of the angles $\phi _b$
such that:
\begin{equation}
\phi _{bn}=0
,\end{equation}
and the angle of the line of nodes in the galactic disc is:

\begin{equation}
\phi _n=\phi _0\pm \pi /2
.\end{equation}

We change the variables of integration
in the expression (\ref{tau2}) to $R$ and $\phi $
(the Jacobian of the transformation is $\frac{\partial \phi _b}{\partial \phi }
\frac{\partial b(R, \phi )}{\partial R}$; we neglect 
$\frac{\partial \phi _b}{\partial R}$ due to the dependence $\theta _0(R)$) 
and obtain

\[
\vec{\tau}(R)=\frac{\rho _{\rm b}v_0^2}{|\sin \theta _0|}
\int _0^{2\pi }\frac{d\phi (1-e_{0Q}(\phi )^2)^{-1/2}}
{1+\sin ^2(\phi _0-\phi)(\sin ^{-2}\theta _0-1)}
\]\[\times
[\sin \phi
\sin \theta _0\vec{i}-\cos \phi \sin \theta _0 \vec{j}+
\cos \theta _0\sin (\phi _0-\phi)\vec{k}]
\]\begin{equation}\times
\frac{\partial b(R, \phi )}{\partial R} b(R, \phi )^2
;\end{equation}
where $b(R,\phi )$ is derived from (\ref{RQ}):

\begin{equation}
b(R, \phi )=\frac{1}{2}R\sqrt{1-e_{0Q}^2(\phi )}
\label{b}
\end{equation}\[
+\sqrt{
\frac{1}{4}R^2(1-e_{0Q}^2(\phi ))+R\ G\ M_{\rm gal}v_0^{-2}(1-e_{0Q}(\phi ))}
.\]

If we make a new change of variables in the integral, substituting
$x=e_{0Q}(\phi )$ for
$\phi $ according to (\ref{e0Q}), and simplify the expression
(some terms cancel because of the antisymmetry in the interval
$(\phi -\phi _0)\in (0, \pi)$ and  $(\phi -\phi _0)\in (\pi, 2\pi)$), we
obtain 

{\small
\begin{equation}
\vec{\tau}(R)=\frac{2\rho _{\rm b}v_0^2\sin \theta _0}
{|\sin \theta _0|\cos ^2\theta _0}
(\sin \phi _0\vec{i}-\cos \phi _0\vec{j})
\label{tau4}
\end{equation}\[ \times
\int _{-\cos \theta _0}
^{\cos \theta _0}dx\ x\frac{\partial b(R,x)}{\partial R} b(R,x)^2
\]\[\times \frac{1}
{\sqrt{1-x^2}\sqrt{1-x^2/\cos ^2\theta _0}[1+(1-x^2/\cos ^2\theta _0)
(\sin ^{-2}\theta _0-1)]}
\]
}

The integral is positive.
The direction of the torque is $\pm (\sin \phi _0\vec{i}-\cos \phi _0\vec{j})$;
the sign is `+' when $\theta _0<0$ and `-' when $\theta _0>0$.
That is, the torque is in the disc and perpendicular to $\vec{v_0}$.

The direction of the infall of a particle flow 
is the same as that produced by the gravitational torque when $\vec{e_0}=
\vec{e_P}$, although the amplitude is rather different. 
Note, for instance, that
the amplitude in (\ref{tau4}) does not depend on the disc density.
Therefore, the effect produced by the infall of the particle beam is
similar to that due to gravitational effects. 

The torques between rings do depend on
disc densities (expression (\ref{2rings}) is proportional to $\sigma (R)
\sigma (S)$). This explains why the disc 
will warp significantly only where its surface density is
low (Ostriker \& Binney 1989) which, in practice, means at its
outer edge. For the inner disc, $\sigma $ is high enough to provide
strong torques between two rings with a small angle $\alpha _{R, S}$ of
separation which can compensate the difference of precession with
respect to the average. However, external rings must separate further
to compensate these differences.

Note that there is no $z$-component of the torque.
Taking into account that the total mass of the ring is increased, this
means that the angular velocity of rotation should decrease and, therefore
the radius of the orbit will be reduced (the increasing mass of the inner 
galaxy will also tend to reduce the radius) and the galaxy will concentrate
further material in the inner regions. We will not study these aspects
further in the present paper.

The limit of low initial velocity ($\frac
{Rv_0^2(1+e_{0Q})}{4G\ M_{\rm gal}}<<1$) gives a proportionality

\begin{equation}
\lim _{\left(\frac{Rv_0^2(1+e_{0Q})}{4G\ M_{\rm gal}}\right)\longrightarrow 0}
\vec{\tau}(R)\propto \rho _{\rm b}\frac{(G\ M_{\rm gal})^{3/2}}{v_0}
R^{1/2}
\label{lim}
\end{equation}\[\times 
(\sin \phi _0 \vec{i}-\cos \phi _0 \vec{j})
.\]
This means that a low velocity beam provide a stronger torque.
This effect can be seen due to the increased curvature of the hyperbolic 
trajectories for low velocities. 
A ring will then accrete flow particles over a wider range of
impact parameters, if the velocity of the beam is lower.
In the extreme case of $v_0=0$ all the particles fall to the centre, so
$R=0$ for all cases with a finite impact parameter and there is no
divergence ($R\propto v_0^2$ from (\ref{RQ})). Although we have
integrated $b$, the impact parameter, from zero to infinity, in a real
case $b$ is limited to a finite value. Expression
(\ref{lim}) means that at lower flow velocities the
disc accretes particles from a greater fraction of the total stream of
particles, and
this is the reason why extra angular momentum is deposited.
Again, as emphasized above, we must note the angular momentum
deposited within a given annulus is due to accretion of particles from
a non-axisymmetric volume of space whose integrated angular momentum
is thus non-zero.

\subsection{Mass accreted by the galaxy}

The total accretion rate due to this infall is

\[
\frac{dM}{dt}=\int _0^{2\pi }d\phi _b\int _{0;\ R<R_{\rm max}}^\infty
db\ b \frac{dm}{dt}=
\rho _{\rm b}v_0\int _0^{2\pi }d\phi \frac{d\phi _b}{d\phi}
\]\[
\left[\int _0^{R_{\rm max}}dR\ b(R,\phi )\frac{\partial b(R,\phi )}
{\partial R}\right]
\]\[
=\frac{-\rho _{\rm b}v_0}{|\sin \theta _0|\cos \theta _0}
\int _{-\cos \theta _0}^{\cos \theta _0}dx b(R_{\rm max},x)^2
\]\begin{equation}\times
\frac{1}
{\sqrt{1-x^2/\cos ^2\theta _0}[1+(1-x^2/\cos ^2\theta _0)
(\sin ^{-2}\theta _0-1)]}
\label{accretion}
.\end{equation}
This rate is positive since the integral is negative
($b(R,-x)>b(R,x)$).

\subsection{Transfer of linear momentum}
\label{.linearmom}

It is important at this stage to consider the transfer of linear momentum. 
While the transfer of angular momentum
produces a integral sign shape (component $m=1$ of the galactic warp), as shown above,
the linear momentum produces a cup-shaped deformation of the disk (component $m=0$ of the galactic warp).
The numbers in the section \S \ref{.MW} will show that
this effect is indeed quite small, and the $m=1$ component is predominant.

The points of equilibrium between the
vertical gravitational force and the vertical force due to transfer of linear
momentum give the deformation of the disk.
The vertical linear momentum due to the accretion of a particle of mass $dm$ is:

\begin{equation}
p_z=dm\ v\sin(\vec{R_Q},\vec{v})\sqrt{1-\cos ^2\theta _0
\sin ^2(\phi _0-\phi)}
\label{pz}
.\end{equation}
The vector $\vec{v}$ is the velocity at the impact point of the disc ($\vec{R_Q}$).
The last two factors account for the projection of the velocity onto the
vertical axis. It is not easy to visualize the origin of these factors; but
they can be understood by reference to Figures \ref{Fig:disc2} and \ref{Fig:orbit}.
The radial and azimuthal linear momentum would produce some distortion
of the orbits within the disc, which is not the subject of the present paper.

From the conservation of the angular momentum, we have:

\begin{equation}
|\vec{J}|=dm\ v_0b=dm\ v\ R_Q\sin(\vec{R_Q},\vec{v})
\label{consJ}
.\end{equation}

From expresions (\ref{pz}) and (\ref{consJ}), the vertical force of a particle of mass $dm$ is:

\begin{equation}
F_z=\frac{dp_z}{dt}=\frac{dm}{dt} 
\frac{v_0b}{R_Q}\sqrt{1-\cos ^2\theta _0
\sin ^2(\phi _0-\phi)}
.\end{equation}

The total vertical force on all the particles of the beam which fall into the ring of radii
$R$ and $R+dR$ is (from now on, $F_z$ stands for the vertical force per unit galactocentric
radial length):

\begin{equation}
F_z(R)dR=\int _0^{2\pi }d\phi _b\int _{0;\ R<R_Q<R+dR}^\infty db\ b \frac{dm}{dt} 
\frac{v_0b}{R_Q}
\end{equation}\[\times
\sqrt{1-\cos ^2\theta _0
\sin ^2(\phi _0-\phi)}
\]\[
=\frac{2\rho _{\rm b}v_0^2dR}{|\sin \theta _0|\cos \theta _0}
\int _{-\cos \theta _0}
^{\cos \theta _0} dx \sqrt{1+x^2-c^2}\frac{\partial b(R,x)}{\partial R}\frac{b(R,x)^2}{R}
\]\[\times
\frac{1}{\sqrt{1-x^2/\cos ^2\theta _0}[1+(1-x^2/\cos ^2\theta _0)
(\sin ^{-2}\theta _0-1)]}
.\]

This acceleration is compensated by the vertical gravitational acceleration, so:

\begin{equation}
\frac{F_z(R)dR}{2\pi \sigma (R)RdR}\approx \frac{GM_{\rm gal}(R)}{R^2}\frac{z}{R}
,\end{equation}
which implies, for a mass derived from the rotation curve, that
\begin{equation}
z\approx \frac{F_z(R)R}{2\pi v_{\rm rot}(R)^2\sigma (R)}
\label{zlinear}
\end{equation}

This monopolar approximation would be exact if the distribution of mass were spheroidal or
elliptical. For the disc, whose contribution is dominant in the torque, the contribution of
the gravitational force is different:

\[
\frac{F_z(R)}{2\pi \sigma (R)R}=Gz\int _0^{R_{\rm max}}ds\ s\ \sigma(s)
\]\[\times
\int_0^{2\pi }
\frac{d\phi}{[s^2+R^2-2sR\cos \phi +z^2]^{3/2}}
\]\begin{equation}
\approx
Gz\int _0^{R_{\rm max}}ds\ s\ \sigma(s)\int_0^{2\pi }
\frac{d\phi}{[s^2+R^2-2sR\cos \phi]^{3/2}}
.\end{equation}
The last approximation is for $z$ small, compared to $(R-s)$; this is a good approach for
large $R$. This is also an approximation in another sense: it does not take
into account the distortion of the disc produced by both $m=0$ and $m=1$ components
of the warp. In any case, most of the mass is in the inner rings which are
not distorted and this effect is small.
Hence, $z$ is inversely proportional to the attraction of the disc even for a non-monopolar
approximation.

The disc contribution is something larger than its monopolar contribution, so
$z$ is lower than (\ref{zlinear}).
This means that the expression (\ref{zlinear})
gives, approximately, a maximum limit for the distortion of the disc, i.e.

\begin{equation}
z\le \frac{F_z(R)R}{2\pi v_{\rm rot}(R)^2\sigma (R)}
\label{zlinear2}
.\end{equation}

This approximation is good enough when the effects of the transfer of linear momentum
are nearly independent of those due to the transmission of angular momentum.
A more accurate model would refine these approximations and calculate both
the $m=0$ and $m=1$ distortions simultaneously. We will see later that the effect of
the transfer of linear momentum is small compared with the transfer of angular
momentum and, therefore, we do not need to make these more refined calculations for this
effect, but only for the integral sign warp, which is dominant.

Is it counterintuitive to have $m=1$ deformation larger than $m=0$ deformation?
We can argue that it is not.
Imagine a galaxy in which practically all the mass is in the centre, so that
the disc has a negligible mass. How big is the countertorque of the
galaxy to compensate the external torque? The answer is: zero (or nearly 
zero), because only the dipole and higher order components of the gravitational
attraction produce a countertorque. The monopolar component does not create
countertorque. In this case, what will be the equilibrium condition in which  
the torque is compensated by the countertorques of the disc? It is an 
``$m=1$ deformation of the disc'' (i.e. a warp) whose amplitude tends to infinity 
(as the mass of the disc tends to zero). In fact the limiting case will not be
at infinity, since we are talking about an angular amplitude. The ring will 
take the angle of incidence of the wind as its maximum limit. 
So what happens to the ``$m=0$ deformation''?
This deformation is nothing like as strongly dependent on the distribution of
the mass and there is a net finite counterforce of the galaxy to compensate 
the external force; the amplitude of the $m=0$ cup-shaped distortion is thus
finite. Therefore, we have a very high amplitude for $m=1$ 
and the amplitude of $m=0$ is much smaller.
If the density of the wind tends to zero, the amplitude of $m=0$ distortion 
tends to zero, but the amplitude of the $m=1$ distorsion tends to take up the   
incidence angle of the wind. 
We conclude that this is a very clear case in which the $m=1$ distortion 
amplitude is much bigger than the $m=0$ amplitude and, therefore, it is possible
to have a $m=1$ deformation larger than the $m=0$ deformation.

\subsection{Direction of the warp}
\label{.direc}

One relevant aspect of the present hypothesis is that it allows us to
relate the direction of the warp
to the direction of the inflow, $\vec{v_0}$, with respect to the
centre of the galaxy.

The torque between two rings, $d\tau _{S}(R)dR\ dS$, is proportional to
$(\sin \phi _0 \vec{i}-\cos \phi _0\vec{j})$.
This can be demonstrated
by use of the equation (\ref{2rings}), or by a simple analogy
with the gravitational torque of one particle (\S \ref{.gravtor}): 
the second ring is the equivalent to an integration of the particle position 
over the second ring azimuthal angle. 
Likewise, the external torque is proportional to 
$(\sin \phi _0 \vec{i}-\cos \phi _0\vec{j})$.
This means that the warp
direction is the same of the projection of $\vec{v_0}$ over the disc,
and the precession is around the axis parallel to $\vec{v_0}$.

The amplitude of the collisional torque, (\ref{tau4}), 
is larger for larger $R$, since $b^2\frac{\partial b}{\partial R}$ 
increases with $R$, so the outer rings have an excess angular
velocity of precession compared to the average. However, the 
outer rings receive a torque from the inner rings,
which is proportional to the collisional torque when $\phi _P=\phi _0$,
where $\vec{e_P}$ is the position of the maximum height above the plane of
both inner and outer rings (i.e. the direction of the warp), 
but must act in the opposite direction to counteract the above excess.
Hence, the outer rings should be oriented towards positive $z$ if 
$\theta _0$ is positive or towards negative $z$ if $\theta _0$ is negative.

\begin{figure}
\begin{center}
\mbox{\epsfig{file=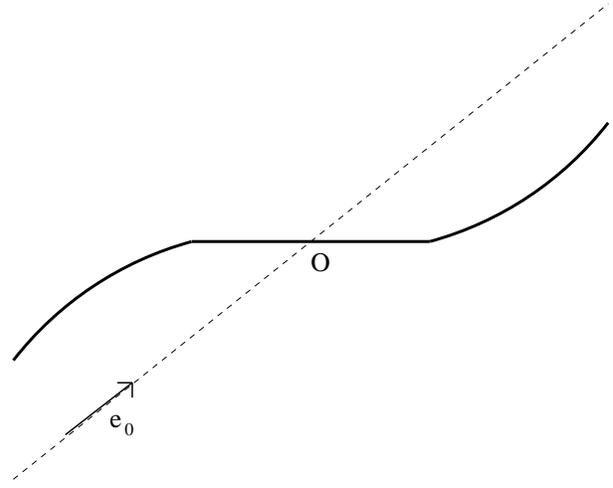,width=8cm}}
%\mbox{\epsfig{file=empty.eps,width=8cm}}
\end{center}
\caption{Graphical representation of the orientation of the warp with
respect to the velocity of incoming flow. The azimuthal angle of the 
warp, which is not plotted, is the same as the azimuthal angle of
$\vec{v_0}$, i.e. $\phi _0$.}
\label{Fig:warpdirec}
\end{figure}

From the above considerations, we deduce that the warp is oriented 
in the direction parallel to the projection of $\vec{e_0}$ in the disc.
The azimuthal angles of the maximum  
and minimum heights ($z$) of the warp are $\phi _0$ and $\phi _0+\pi $ 
respectively if $\theta _0>0$. If $\theta _0<0$, the maximum height is at
azimuth $\phi _0+\pi$ while the minimum is at $\phi _0$.
Figure \ref{Fig:warpdirec} shows a graphical representation of
these orientations in the plane which contains the vector
$\vec{v_0}$ perpendicular to the disc. 
Note that the orientation of the warp does
not change if the velocity is $-\vec{v_0}$ instead of $\vec{v_0}$.

\section{Warp in a typical spiral galaxy due to intergalactic accretion
flows}
\label{.MW}

The proposed hypothesis of formation of warps can be compared
with our observational knowledge about galaxies and the intergalactic medium.
We must bear in mind that the calculations developed in this paper
are based on fairly crude approximations. We have considered particle
trajectories as hyperbolic, neglecting the gravitational effects
of the extended disc, we have considered a collimated infinite beam, constant
density of the intergalactic flow, etc. Therefore,
we should not expect the theory to reproduce exactly all the fine details
of a warp. However, we will show that our model
reproduces quite well the observed warps using a standard model
of the disc of a typical spiral galaxy (for instance, the Milky Way)
and with entirely reasonable parameters for the intergalactic
flow. Our restricted goal here is
to offer a general model for warps (a warp amplitude $\alpha (R)$
which is close to zero out to some specific $R$, but rises rapidly at higher
increasing $R$), and to test it using the parameters of a typical 
spiral galaxy such as the Milky Way.

\subsection{Amplitude of the warp}

The solution of (\ref{domegat}) by means of the numerical method explained
in appendix \ref{.alpha0} gives $\alpha (R)$ for the galactic warp.
It will depend on the adopted galactic model.
Our purpose is to demonstrate that the order of magnitude
of the amplitude of the warp is approximately the observed amplitude, so
we use only one model. Varying the parameters of the disc
will, of course, lead to variable results but these would not 
be quantitatively very different.
The surface density of the disc we adopt is given by (\ref{sigmaMW}) and 
the rotation velocity is given by (\ref{rotvelMW}), both corresponding to the Milky
Way.

In the linear regime, for a low amplitude of the warp and low velocity
$v_0$, in the limits (\ref{lim}) and (\ref{lim2rings}), the proportionality
followed is $\alpha (R)\propto \omega _p \propto \frac{\rho _{\rm b}
M_{\rm gal}^{3/2}}{v_0}$,
in which the dependence of $\alpha (R)$ on $R$ is not specified.
Specifically, for the adopted disc model and $|\theta _0|=45^\circ$,
we obtain in this linear regime that the angle of the warp is
($\Omega _b=0.02\ {\rm h^{-2}}$; Schramm \& Turner 1998)

\[
\alpha (2R_0)\approx 5.2\times 10^{-34}\frac{\rho _{\rm b}({\rm kg/m}^3)
M_{\rm gal}({\rm kg})^{3/2}}{v_0({\rm m/s})}\ {\rm rad}
\]\begin{equation}=
1.7\times 10^{-4}\frac{\rho _{\rm b}}{\Omega _{\rm b}\rho _{\rm
crit}}
\left(\frac{M_{\rm gal}}{10^{11}\ {\rm M_\odot}}\right)^{3/2}
\frac{100\ {\rm Km/s}}{v_0}\ {\rm rad}
,\end{equation}
and the precession angular velocity

\begin{equation}
\omega _p\approx 7.7\times 10^{-21}
\frac{\rho _{\rm b}}{\Omega _{\rm b}\rho _{\rm
crit}}
\left(\frac{M_{\rm gal}}{10^{11}\ {\rm M_\odot}}\right)^{3/2}
\frac{100\ {\rm Km/s}}{v_0}\ {\rm rad/s}
.\end{equation}

Burton (1988) gives an observational value of $\alpha (2R_0)\approx 0.14$ rad
in our Galaxy which, if we take the value at $2R_0$ to normalize the amplitude
of the warp, leads to 

\begin{equation}
\rho _{\rm b}\approx 820\ \Omega _{\rm b}\rho _{\rm crit}
\left(\frac{v_0}{100\ {\rm Km/s}}\right)
\left(\frac{M_{\rm gal}}{10^{11}\ {\rm M_\odot}}\right)^{-3/2} 
,\end{equation}
\begin{equation}
\omega _p\approx 6.2\times 10^{-18}\ {\rm rad/s}=1\ {\rm cycle\ in\ 32\ Gyr}
.\end{equation}
The precession is far too slow to be significant, and much slower than
the rotational velocity of the Galaxy.

We adopt a total mass of the Galaxy of 
$M_{\rm gal}=2\times 10^{11}$ M$_{\odot}$ (Honma \& Sofue 1996)
to calculate the curvature of the hyperbolic trajectories. 
This estimate includes bulge, disc and spheroidal component masses.
Hence,

\begin{equation}
\rho _{\rm b}\approx 290\ \Omega _{\rm b}\rho _{\rm crit}
\left(\frac{v_0}{100\ {\rm Km/s}}\right)
\end{equation}

The only free parameters with observational uncertainties are
the mean density of the intergalactic inflow and its velocity which
are related by the last equation. Assuming a typical
velocity, in galactocentric coordinates, of $v_0\sim 100$ km/s (the limit
of low velocity adopted in (\ref{lim}) is valid if $R$ is much less
than 300 or 400 kpc, which is certainly the case since we take
$R\le 16$ kpc), we find:

\begin{equation}
\rho _{\rm b}\sim 290\ \Omega _{\rm b}\rho _{\rm crit}=
1.1\times 10^{-25} \ {\rm kg/m^3}
,\end{equation}
which is equivalent to
\begin{equation}
n_{HI}\sim 6\times 10^{-5} f\ {\rm cm}^{-3}
\label{densHI}
,\end{equation}
where $f$ is the fraction of HI in the total baryonic mass of the wind.
This value coincides with the most probable value estimated by
Kahn \& Woltjer (1959) for the mean density of intergalactic matter
yielding dynamical stability for the Local Group of galaxies.
L\'opez-Corredoira et al. (1999) also calculated a total intergalactic
mass in the Local Group around $2\times 10^{12}$ M$_\odot$, which in a
volume of $\sim 1$ Mpc$^3$ gives a mean density around $10^{-25}$ kg/m$^3$.

\begin{figure}
\begin{center}
\mbox{\epsfig{file=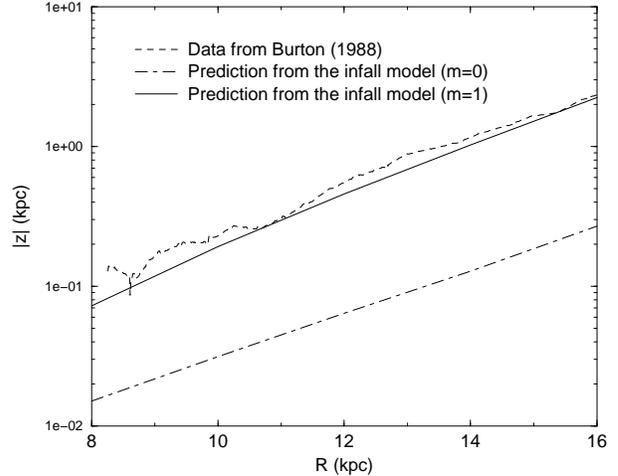,width=8cm}}
%\mbox{\epsfig{file=empty.eps,width=8cm}}
\end{center}
\caption{Milky Way warp maximum height 
as a function of the radius in the hypothesis
of continous accretion of intergalactic matter with $\rho _{\rm b}=1.1\times
10^{-25}$ kg/m$^3$, $|\theta _0|=45^\circ$ ($m=1$, solid line). 
For comparison, a maximum limit of the effects of the linear momentum 
increment over the rings are shown ($m=0$, dot-dashed line).
Dashed line stands for observational data of the northern warp
(Burton 1988) rescaled to $R_0=7.9$ kpc.}
\label{Fig:warpMW1}
\end{figure}

The shape of $\alpha (R)\approx |z|/R$ for these values would be that given in
Fig. \ref{Fig:warpMW1}. The prediction of this model agrees with many of the
features of the observed warp (Burton 1988; 1992): it predicts a flat disc which
is not significantly warped for values of $R$ less than $\sim 1.3\ R_0=10$ kpc, 
and a warp whose amplitude increases rapidly at larger radii. 
The shape of the warp depends
on the detailed radial variation of $\sigma (R)$ and we have used a
rough exponential estimate of this dependence, which is really more complex
than a simple exponential function with a constant scale length.
In any case, it is not at this stage worth using a more realistic model since
our approach is only an approximation, given the simplifications
we have employed in the analytical calculation of the
torque. The functional shape of $\alpha (R)$ resembles that
of the observational data, within the values of $R$ less than $\sim 16$ kpc 
from the centre. Non-linear effects have a significant influence
at larger radii as
well as departures from the simple ring model of the disc.
Nevertheless, we should note that the dependence of $\vec{\tau }_{\rm ext}$ 
on $R$ does not affect too much the form of $\alpha (R)$
and other external torques may also produce similar shapes, as we have
seen in Figures \ref{Fig:alfagrav} and \ref{Fig:alfamag}.

\begin{figure}
\begin{center}
\mbox{\epsfig{file=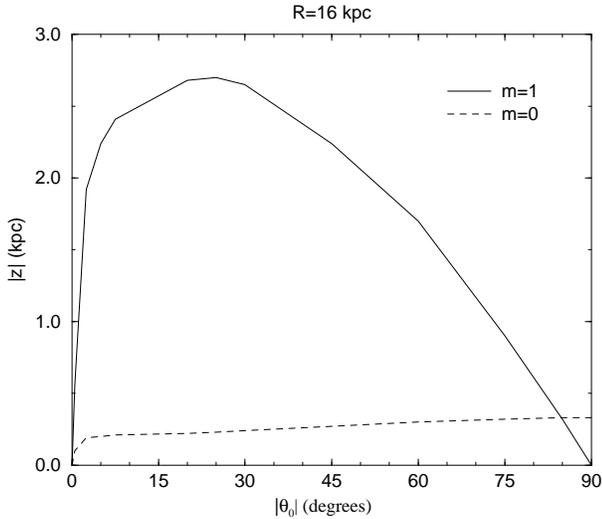,width=8cm}}
%\mbox{\epsfig{file=empty.eps,width=8cm}}
\end{center}
\caption{Milky Way warp maximum height 
as a function of the inclination of the incoming flow with respect to
the galactic plane ($\theta _0$) in the hypothesis
of continous accretion of intergalactic matter: 
$m=1$ (integral-sign warp due to a torque), 
solid line; $m=0$ (cup-shaped distortion due to a force), dashed line.}
\label{Fig:warp_angle}
\end{figure}

The maximum distortion of the disc due to the transfer of linear momentum (\S \ref{.linearmom})
is calculated using the expression (\ref{zlinear2}). The results are also plotted in
Fig. \ref{Fig:warpMW1}. In this case, $z$ is always positive or always negative,
a cup shape rather than an integral sign shape. From the numbers plotted in this figure,
it can be concluded that this
distortion is small compared to that produced by the torque, i.e. the integral
sign shape will be predominant in the galactic disc.
We think the predominance of torque effects is due to the
smaller mass concentrated in the disc in comparison to the total mass
of the Galaxy, which is mostly responsible of the counter-torques 
which reduce the heightness of the warp $m=1$, while the whole mass 
of the Galaxy produces the counter-forces which reduce the distortion $m=0$.
In some galaxies, a mixture of cup-shape and integral sign shape could be produced
by this mechanism. This would be a possible explanation for the asymmetries between
the two sides of the warp, even for our own Galaxy.
It will depend, among other factors, on the angle $\theta _0$. For $\theta _0=90^\circ $,
the transfer of vertical linear momentum would be maximum while the torque would be
zero. Therefore, there is a probability, although small, of finding 
cup-shaped distortions of galactic discs rather than integral-sign warps.
In Fig. \ref{Fig:warp_angle}, we show that the cup-shaped distortion is
predominant for $|\theta _0|>\sim 85^\circ$, which translates to a probability
that a cup-shape is dominant of $<\sim 4\times 10^{-3}$, for spiral galaxies like
the Milky Way. It would be of interest to make a statistical search over a major sample of warped galaxies
to see how many galaxies present appreciable cup-shaped distortion.
The study of the fraction of galaxies which present 
irregulatities within the integral sign warp would also give some clue about
the validity of this hypothesis.

Near the limit $\theta _0=0$ the amplitude of the warp goes to zero quickly.
Although, the figure Fig. \ref{Fig:warp_angle} shows apparently a vertical tangent
in the limit $\theta _0=0$, that is not the case: there is a horizontal tangent,
as the limit of (\ref{tau4}) when $\theta _0$ is very small is proportional
to $\theta _0^2$. However, this limit manifests for very small angles 
and this is the reason for the apparent vertical tangent in \ref{Fig:warp_angle}.
Indeed, the values are: $|z|(\theta _0=10^\circ )=2.45$ kpc, $|z|(\theta _0=5^\circ )=2.24$ kpc,
$|z|(\theta _0=2^\circ )=1.79$ kpc, $|z|(\theta _0=1^\circ )=1.24$ kpc, $|z|(\theta _0=0.5^\circ )=0.59$ kpc,
$|z|(\theta _0=0.25^\circ )=0.19$ kpc, $|z|(\theta _0=0.1^\circ )=0.034$ kpc,..., $z(\theta _0=0)=0$ kpc.

We can add some further comments about the stellar warp.
In the Milky Way, the OB stars follow the gas (Porcel \& Battaner 1995). 
But it is not just the young population, recently formed from the warped gas,
which shows a stellar warp, but the
whole population of stars (Porcel et al. 1997; Dehnen 1998). The whole
population of stars of the Milky Way projected onto the sky appears
less deviated from the plane than the projected gas ,
but this does not necessarily represent a difference between 
amplitudes of the gaseous and stellar warps.
Rather, it may well be due to the cut-off of the stellar population
at a smaller Galactocentric radius than the gas, around 15 kpc
(Porcel et al. 1997). 
The stellar disc is clearly warped, perhaps somewhat less
than the gaseous disc although not obviously so. As pointed out in \S
\ref{.1part}, if the stellar disc were
demonstrated to be less warped than the gas warp, it would be evidence
supporting either this theory or the theory
of the intergalactic magnetic field as the generator of the warp
(Porcel et al. 1997). In fact,
this would be predicted by a theory in
which the external torque directly affects the gas disc and
only indirectly the stellar disc.

\subsection{Possible scenarios}

Possible scenarios for the intergalactic flow postulated here are:

\begin{enumerate}

\item
The galaxy is passing through a continuous intergalactic medium,
i.e. the velocity of the flow is due to the relative motion of the
galaxy with respect to the rest frame of the intergalactic medium.
In this case, the flow may be well approximated by a beam of infinite
extent. This may be a representative scenario for most galaxies, 
and could explain why most spiral galaxies are warped.
Asymmetries in the warp could also be explained in terms of
a transfer of both linear and angular momentum, although the 
cases when cup-shaped distortions dominate has low probability.
We think that this first item is the most plausible explanation.

\item The galaxy has an interacting companion and some exchange
of material is produced due to tidal effects. The stream of material goes
from the companion to the main galaxy, and this flow could also generate
a warp. The direction
of the wind would not be constant, because the companion is orbiting 
around the main galaxy so the warp would not be steady.
This scenario would give
asymmetries in the warp, apart from that coming from the $m=0$ component,
which would be due to the non-symmetric form of the infalling
material, with finite impact parameter, or clouds which intersect 
the centre of the galaxy in a non-axisymmetric way.
Reshetnikov \& Combes (1998) find some correlation between
the frequency of warps and the interaction with other galaxies.
An accretion inflow due to the exchange of material with a companion might appear
to be undermined by the observations of warped galaxies which are apparently
isolated. However, this phenomenon is itself being challenged by
the discovery of some companions
to galaxies which had been considered as classical examples of isolated galaxies
(for instance, NGC 5907; Shang et al. 1998).

\end{enumerate}

While the present paper was in the refereeing process, a work by
Garc\'\i a-Ruiz (2001) has been published. This is an interesting
work which analyzes in detail 26 edge-on galaxies in radio and optical.
Garc\'\i a-Ruiz (2001) finds that 20 galaxies are
warped, two of them present U-warp, and 7 with a warp of only one side.
This asymmetric cases can be explained by the present theory as a combination
of $m=0$ and $m=1$ distortion and it is interesting to note
that up to now there is not any alternative explanation for the U-warps
as well as the asymmetric warps.
He also find that the frequency of warps and its amplitude
is dependent on environment. It is even more interesting to 
note that the most isolated galaxies are more frequently warped
(although with less amplitude).
{\it ``While this suggests that tidal interaction plays 
a role in warping, it seems
likely that there are other effects at work that cause even quite 
isolated galaxies to warp.''} (Garc\'\i a-Ruiz 2001).
It seems clear that warping is due to something related with
the environment rather that the intrinsic propierties of the galaxies,
and something which is not related with the proximity of other galaxies.
The accretion of intergalactic matter onto the disc seems a good
candidate to explain these observational facts.

The density of the
intergalactic medium is very low on average, making it
difficult to detect. However, it could well be
that the High Velocity Clouds (HVCs; see the reviews in 
Wakker \& van Woerden 1997; Wakker et al. 1999b) are part or all of this material 
falling towards the Galactic disc from all directions.
If these do produce a net torque it must be
because there is a net galactocentric average velocity of the complete
set of HVCs, including the Magellanic Stream, the complexes and
the HVCs associated with more distant clouds infalling towards the 
Local Group barycenter (Blitz et al. 1999;
Braun \& Burton 1999; L\'opez-Corredoira et al. 1999).
The tidal Stream from the Saggitarius dwarf galaxy (Ibata et al. 2001) might also
take part.
An average hydrogen density for the flux of 
$n_{HI}\sim 6\times 10^{-5}f$ cm$^{-3}$
is in good agreement with the average density of an HVC, 
$\langle n_{HI}\rangle \sim 10^{-4}$--10$^{-1}$ cm$^{-3}$ (Blitz et al. 1999; 
Wakker et al. 1999a), including the value of $f$ which may be as low as 
0.02 (Braun \& Burton 2000) if the matter of the flow is baryonic. 
The mean density in the intergalactic
clouds may be enough to create the Galactic warp. 
If the density of infalling matter
were equal to that of an individual HVC, we would have
a higher density than that required, by several orders of magnitude. 
However, we know that the intracluster medium
is not filled by HVCs; these represent a low volume high density
fraction; averaging over the complete medium with plausible filling
factor can yield a net
hydrogen density around $n_{HI}\sim 6\times 10^{-5}f$ cm$^{-3}$. 

The degree of clumpiness of the intergalactic medium is not
well known. However, it is not of importance whether the flow is continuous or discretized in clouds.
The warp will be produced by the average infall. Short-term
fluctuations of the infalling density do not appreciably affect
the warp since the forces responsible for it have a very low amplitude
and require a long time to produce or distort the warp.

Using expression (\ref{accretion}), we can derive a
total accretion rate of the Galactic disc out to $R_{\rm max}=15$ kpc
of $\sim 1$ M$_\odot$/yr for this density. This 
turns out to be of the order of the accretion rate required to resolve 
the G-dwarf problem in our Galaxy
as well as explaining a number of phenomena
of chemical evolution
which require the long-term infall of low metallicity gas 
(L\'opez-Corredoira et al. 1999; Wakker et al. 1999a).
The infall of 1 solar mass per year is enough to produce the
warp because: 1) the external disc has a very low density
and small forces produce considerable accelerations; 2) the acceleration
may in fact be very small in amplitude, but the period of time to produce
the warp is large enough (order of Gyr), so this gives time to distort
the galactic disc (in 1 Gyr, $10^9$ solar masses are accreted which is 
a significant quantity of accreted mass).

This hypothesis could, in a general way, explain 
the possible alignments of the 
different warps of neighbouring galaxies (Battaner et al. 1991) if the flow
velocity is similar around such galaxies within the same zone of the
intergalactic medium.

Whatever the structure and composition of the intergalactic medium, 
it is clear that
intergalactic space is by no means empty, and the accretion of this
material by galaxies is likely to have been continuous
during their lifetimes. 
The effects of this accretion can be detected in their chemical evolution
as well as in their structure, as pointed out above.

\section{The effects of a very massive halo}
\label{.halo}

Although we already discussed the effects of the halo, 
we are going to clarify
the use of the halo here as well as the effects which
a very massive halo could produce.

We may infer from the present calculations that we have found
a mechanism which could explain both qualitatively and quantitatively
the generation of warps in normal spiral galaxies,
and this mechanism is the interaction between the disc
and the infalling matter as well as the interaction of the disc with itself
(rings interacting with other rings).
The dark halo is included although it is not explicit in some calculations
of this paper, but its effect is not very important for a rough
calculation in which we are interested in the order of magnitude. 
In fact, our calculations are not rough but almost exact, but the roughness of the numbers
obtained is due the uncertainties in the different parameters.

Briefly, these are the effects which are treated here, and
others which have not been treated but require treatment in future papers:

\begin{itemize}

\item The dark halo is present in the total mass of the Galaxy: $M_{\rm gal}$.
The assumed value of the Milky Way mass of $2\times 10^{11}$ M$_\odot$
within a radius $R\sim 20$ kpc from the centre
includes the halo, some of whose matter is not visible.
The effects of the halo are also implicitly included in the
calculation of $M_{\rm gal}(R)$ as a function of $v_{\rm rot}(R)$
from the rotation curves.

\item The interaction of the halo with the warp is not considered, but
we have made estimates to show that for this case the effect is small compared to 
the countertorque of the disc. Therefore, 
we feel that the exclusion of halo effects in the countertorques
is a fair approximation if we are interested only to find the order of magnitude of the 
warp amplitude.
The countertorques of the halo are less important than those of 
the disc because the halo is much closer to sphericity.
We showed in \S \ref{.warptor} that 
the contribution of the halo countertorque for $R<16$ kpc is less than $\sim 40$\% of the
disc contribution. The uncertainties in the halo mass distribution lead
to an uncertainty similar to this value and we did not feel that much would
be gained by adding the extra complexity. We know that
the warp would be reduced by this effect, but this reduction is not the
dominant term.

\item A very massive and very extended halo would introduce
some variations in the numbers we derived although the mechanism would
work qualitatively in a similar way. First, if it is very massive
$M_{\rm gal}$ would be larger and the amplitude of the warp would
be correspondingly larger. We have shown that the amplitude of the warp is
proportional to $M_{\rm gal}^{3/2}$. This, however, was on the
assumption that most of the mass is concentrated within a radius less
than $\sim 20$ kpc. If the halo is very extended, this approximation
is not appropriate. We should then need to consider the mass distribution of
the halo instead of assuming a point mass to calculate the infall
velocity of the clouds. The problem would be much more complex because
we would need a gravitational potential different from $\phi \propto
1/r$ and the trajectories would not be hyperbolae. Indeed, the mechanism
we propose to form warps works with any potential, but the calculations
are of course much easier for a $\phi \propto 1/r$ law. Since we 
are interested in proposing a new mechanism and showing how it works,
we think that these complexities should be left for a future paper.
At present, we can say that a very massive halo would increase
the amplitude of the warp, not proportionally to $M_{\rm gal}^{3/2}$
but some other power with exponent less than 3/2. This is because the infall velocity
of the clouds is increased as well as the curvature of the trajectories.
It is important to note that a very extended halo would not differ
from a halo constrained within $R<20$ kpc in the countertorques produced on the warp
because only the mass in the ellipsoids internal to the radius of the warp
produce gravitational torque, assuming there is a constant ellipticity halo. 
To summarize, with a very massive halo our
mechanism does operate, even in fact more effectively 
because it would require a lower intergalactic density to produce the same warp
amplitude. 

\item If the halo axis were displaced with respect the disc axis the disc would be
pinched by the halo. This was the case studied by Ostriker \& Binney (1989),
Binney et al. (1998) or Jiang \& Binney (1999) and is indeed a
mechanism which will give rise to warps too.
The scenario presented by these authors is different from the scenario proposed
here. They assume the infall of the intergalactic matter onto
the halo rather than the disc, and this produces motion of the halo with
respect to the disc. We do not seek to challenge this in the present paper,
rather to present a possible alternative, which might be complementary.
The compatibility of the two mechanisms certainly merits effort to study further.
There is no doubt that accretion into a halo which gives an offset in the rotation axis
from the disc axis will give rise to a warp; if one accepted the possibility
of a low density halo accreting intergalactic matter, one should recognize the
possibility of giving rise to warps by this mechanism. We have shown here how
accretion directly onto the disc can also yield a warp, with parameters in the
observed range. At the present stage of understanding the problem, we think
that either or both mechanisms can act to produce warps if one accepted that
the accretion of matter by the disc or the halo have the same plausibility.
Our opinion is that our mechanism is preferable since the accretion of matter
by the disc is more plausible than the accretion by the halo but this point
is open for discusion.

\end{itemize}

\section{Conclusions}

We propose that galactic warps are produced by the reorientation
of the galactic disc structure in order to compensate the differential
precession due to a torque generated by an external force.
The external force might be the gravitational interaction with a
satellite, but in the case of the Milky Way, with the Magellanic Clouds
as the satellites, there is not enough mass close enough 
to provide the observed amplitude
in the warp. Magnetic forces could also produce the warp, but the
intergalactic field would then be of the order of $\mu $G.

A simple model of an intergalactic accretion flow 
which intersects a galactic disc 
(or, equivalently, considering the galaxy as moving through the intergalactic 
medium) can explain the existence of warps in the galaxy
if the mean density of baryonic matter in 
the medium is around $10^{-25}$ kg/m$^3$ and the 
infall velocity at large distance is $\sim 100$ km/s.
This hypothetical low density flow is a very reasonable physical assumption
and would explain why
most spiral galaxies are warped.

Accretion due to such a flow is in good accord with the observations
of the chemical evolution of the Milky Way by contributing
$\sim 1$ M$_\odot$/yr of low metallicity gas to the disc.
The High Velocity Clouds, which are presumably the
accretable material in the Local Group galaxies (Blitz et al. 1999;
Braun \& Burton 1999; L\'opez-Corredoira et al. 1999;
Wakker et al. 1999a) are candidates for a significant fraction of the material 
which fills intergalactic space and is accreted.
No massive halo is necessary nor high values of magnetic fields
are necessary, although the presence of these elements would not
modify qualitatively the present conclusions. 
No calculations in the framework of accretion onto the disc
are given for a very massive halo but its effect
would not affect qualitatively the present mechanism, as it is explained in
\S \ref{.halo}. Models with a very 
massive halo could be generated and different numerical results 
would be obtained depending on the parameters of the halo, although
no major qualitative changes are expected to the model presented here
since the mechanism of formation of warps is dominated by the interaction with
the disc; only quantitative changes would come from the increasing
velocity and trajectory curvature of the infalling material which
would increase the amplitude of the warp, i.e. the same amplitude
of the warp would be 
obtained with a density even lower than $10^{-25}$ kg/m$^3$.

Several mechanisms can generate
warps but, among them, accretion of an intergalactic flow seems
to offer a very plausible scenario: it is quantitatively 
consistent with many observations
and works independently of other ingredients of galaxies
and their structure.

{\bf Acknowledgements:}
We particularly appreciate the comments of the referees---E. Battaner and the 
anonymous referee---and of J. J. Binney and I. Shlosman, whose detailed 
questions have helped us to explain more clearly some important technical 
points. This work has been supported by grant PB97-0219 of the Spanish DGES.

\begin{appendix}

\section{Numerical calculation of $\alpha (R)$}
\label{.alpha0}

The solution of $\frac{d\omega _p(R)}{dR}[\alpha (R)]=0$, from
(\ref{domegat}),
can be achieved by means of a numerical calculation with $N$ discreet values
of $R$, i.e. 

\begin{equation}
\vec{H}(\vec{\alpha })=\vec{0}
,\end{equation}
\[
\vec{\alpha }=(\alpha _1,...,\alpha _N);\ \vec{H}=(H_1,...,H_N)
,\]
\[
\alpha _i\equiv \alpha (R_i);\ H_i\equiv \frac{d\omega _p}{dR}(R_i)
.\]

Newton-Raphson's iterative method is appropriate for this kind of
numerical calculations. The iteration $k+1$ is given by

\begin{equation}
\vec{\alpha }^{k+1}=\vec{\alpha }^k-{\cal W}^{-1}(\vec{\alpha }^k)
\vec{H}(\vec{\alpha }^k)
,\end{equation} 
\[
{\cal W}_=\left(
\begin{array}{lll}
W_{11} & ... & W_{1N} \\
... & ... & ... \\
W_{N1} & ... & W_{NN}
\end{array}
\right)
\ \ ; W_{ij}=\frac{\partial H_i}{\partial \alpha _j}
\]

By means of this application, we can calculate $\alpha (R)$. The first
iteration is taken with $\alpha (R)=|\theta _0|$.

\end{appendix} 

\end{document}